\documentclass[12pt]{article}
\usepackage[utf8]{inputenc}%
\usepackage[T1]{fontenc}
\usepackage{amsmath}%
\usepackage{amsfonts}%
\usepackage{amsthm}%
\usepackage{booktabs}
\usepackage{tabularx}
\usepackage{caption}
\usepackage{adjustbox}
\usepackage{subcaption}
\usepackage{multirow}
\usepackage{graphicx}

\theoremstyle{definition}

\theoremstyle{remark}

\usepackage[margin=1.2in]{geometry}
\usepackage[onehalfspacing]{setspace}
\usepackage{xurl}
\usepackage{mathtools}

\usepackage{bbm} 

\usepackage{comment}
\usepackage[title]{appendix}
\usepackage{etoolbox} 
\AtBeginEnvironment{appendices}{\crefalias{section}{appendix}}

\usepackage[
bibencoding=utf8,
maxbibnames=3, 
minbibnames=1, 
backend=biber,%
sortlocale=en_US,%
style=apa,
uniquename=false,%
url=true,%
sortcites=false,
doi=true,%
eprint=true%
]{biblatex}
\usepackage{changepage}

\usepackage{xcolor}%
\definecolor{webbrown}{rgb}{.6,0,0}%
\usepackage{hyperref} %
\hypersetup{%
  breaklinks = true,%
  colorlinks = true,%
  anchorcolor = webbrown,%
  citecolor = webbrown,%
  filecolor = webbrown,%
  linkcolor = webbrown,%
  menucolor = webbrown,%
  urlcolor= webbrown,%
  citebordercolor= 1 0 0,%
  menubordercolor=1 0 0,%
  urlbordercolor=1 0 0,%
  runbordercolor=1 0 0,%
  pdftitle={Contagion From a Bank Run},
  pdfauthor={Choi, Goldsmith-Pinkham, Yorulmazer}
}

\usepackage{cleveref}
\crefname{appsec}{appendix}{appendices}
\crefname{appsubsec}{appendix}{appendices}
\crefname{assumption}{assumption}{assumptions}

\usepackage[nolist]{acronym}
\begin{acronym}
  \acro{SVB}{Silicon Valley Bank}%
  \acro{GFC}{Global Financial Crisis}%
  \acro{LCR}{Liquidity Coverage Ratio}
\end{acronym}

\usepackage{accents}

\newcolumntype{Y}{>{\centering\arraybackslash}X}

\providecommand{\keywords}[1]
{
  \small	
  \noindent\textbf{\textit{Keywords:}} #1
}
\providecommand{\JEL}[1]
{
  \small	
  \noindent\textbf{\textit{JEL Codes:}} #1
}

\usepackage{float}
\title{Contagion Effects of the Silicon Valley Bank Run\thanks{\noindent Choi:
\href{mailto:dong.choi@snu.ac.kr}{dong.choi@snu.ac.kr}, Goldsmith-Pinkham:
    \href{mailto:paul.goldsmith-pinkham@yale.edu}{paul.goldsmith-pinkham@yale.edu}, 
    and Yorulmazer:
    \href{mailto:TYORULMAZER@ku.edu.tr}{tyorulmazer@ku.edu.tr}. We thank Viral Acharya, Thomas Eisenbach, Douglas Gale, Samuel Hanson, Stephen Karolyi, Tyler Muir, Jonathan Parker, Adi Sunderam, Eric Zwick and seminar participants at the OCC for helpful comments.}}%
\author{
    Dong Beom Choi\\
    Seoul National University\\
    \and
  Paul Goldsmith-Pinkham\\
  Yale University and NBER
  \and
  Tanju Yorulmazer\\
  Koc University
}
\date{May 2024}

\begin{document}
\maketitle
\thispagestyle{empty} 
\setcounter{page}{0}
\begin{abstract}
This paper analyzes the contagion effects associated with the failure of Silicon Valley Bank (SVB) and identifies bank-specific vulnerabilities contributing to the subsequent declines in banks' stock returns. We find that uninsured deposits, unrealized losses in held-to-maturity securities, bank size, and cash holdings had a significant impact, while better-quality assets or holdings of liquid securities did not help mitigate the negative spillovers. Interestingly, banks whose stocks performed worse post-SVB also experienced lower returns in the previous year, following Federal Reserve interest rate hikes. Stock investors appeared to anticipate risks associated with uninsured deposit reliance, but did not foresee the realization of implied losses. While mid-sized banks experienced particular stress immediately after the SVB failure, over time negative spillovers became widespread except for the largest banks. 
\end{abstract}


\JEL{G01, G21, G14, G28, E58, E43}

\clearpage

\section{Introduction}

On March 10th, 2023, \ac{SVB} failed after a bank run. It was the sixteenth largest bank in the United States as of the end of 2022, with \$209 billion in assets, which marked the largest bank failure since the \ac{GFC} of 2007-2009 and the second-largest bank failure in U.S. history, second only to Washington Mutual.\footnote{See \textcite{acharya2023svb} for an extensive discussion and analysis of the SVB episode.} As a response,  a ``systemic risk exception'' was invoked by the policymakers, leading to a blanket guarantee on all deposits at \ac{SVB}.\footnote{The joint statement by the Secretary of the Treasury Janet Yellen, Federal Reserve Chairman Jerome Powell, and FDIC Chairman Martin Gruenberg on March 12th stated that all deposits at \ac{SVB}, insured and uninsured, would be fully protected.} The failure of SVB, the subsequent bank failures, and the policy responses, such as the blanket deposit guarantee and the new central bank lending facility, underscore the continued significance of bank runs and contagion as important features of the global financial system, even after the implementation of regulatory reforms following the \ac{GFC}. 

Financial contagion can propagate through various channels. These include direct exposures via interbank linkages, information spillovers where difficulties of one bank can be interpreted as a negative signal for others, and disorderly liquidations of illiquid assets. Identifying banks that are affected by contagion during a financial crisis can help delineate the underlying mechanisms driving systemic risk. Market responses, such as measured by changes in stock prices, provide valuable information for quantifying the impact of contagion on other banks. This study adopts the approach of \textcite{GoldsmithYorulmazer2010}, which analyzed contagion associated with the failure of Northern Rock in the United Kingdom during the \ac{GFC}. Specifically, we examine the stock price reactions of banks following the switch to tighter monetary policy and the subsequent failure of \ac{SVB} to characterize which specific bank characteristics played a significant role in driving the spillovers.

In response to historic levels of U.S. inflation following the COVID-19 pandemic, the Federal Reserve began to increase its policy rate on March 17, 2022.  While rate hikes were initially expected to benefit the banking sector by allowing higher net interest margins for new loans (\textcite{drechsler2017deposits}), they also introduced vulnerabilities associated with duration mismatch for existing fixed-rate assets.

\ac{SVB} was particularly exposed to duration risk due to a majority of its investments in highly-rated government bonds with very long maturities. By the end of 2022, \ac{SVB}'s bond portfolio consisted of \$91.3 billion in held-to-maturity (HTM) and \$26.1 billion in available-for-sale (AFS) assets, with mark-to-market accounting unrealized losses  for its HTM securities exceeding \$15 billion.\footnote{``How crazy was Silicon Valley Bank's zero-hedge strategy?'' \textit{Financial Times}, March 17, 2023.} However, these losses were considered ``potential'' and would only materialize if the bank was forced to sell the assets during liquidity stress events such as heavy deposit withdrawals. Amidst mounting concerns, \ac{SVB} experienced a steady decline in deposits over the four quarters leading up to its eventual failure. The pace of withdrawals intensified during the week of March 6th, 2023, culminating in an abrupt outflow of \$42 billion on a single day, March 9th. The run on SVB triggered its failure and subsequent takeover by FDIC on March 10th (\textcite{jiang2023}).

The run on \ac{SVB} triggered significant spillover effects on the U.S. banking system. One crucial factor that contributed to the spillover effects was the concern over unrealized losses in highly liquid securities. Unlike typical bank run scenarios, where liquid assets act as buffers, the realization of implied losses on the liquid securities due to funding outflows proved problematic. Having more liquid securities did not help mitigate contagion; only cash holdings were effective.  Additionally, reliance on uninsured deposits exacerbated financial distress due to their potential for rapid withdrawal. 

The correlation of bank size with spillover effects was also noteworthy. Mid-sized banks, particularly those with assets in the \$50 billion to \$250 billion range referred to as ``super-regional'' banks, experienced heightened stress immediately after the run on \ac{SVB}. These banks had previously been designated as systemically important by the regulators but faced relaxed regulations after the 2018 regulatory rollback. Investors initially exhibited greater concerns with banks of a similar size to \ac{SVB}, potentially due to regulatory differences. Over the next months, however, the negative effects spread to other banks as concerns about the entire banking system became more widespread. Surprisingly, very large banks, particularly those with assets over \$1 trillion, outperformed the rest of the system. This may be attributed to implicit too-big-to-fail (TBTF) guarantees and consequent deposit inflows, since these banks were perceived as safe havens during the systemic instability (\textcite{caglio2023}, \textcite{LuckPlosserYounger2023}, \textcite{BaronSchularickZimmermann2023}).

Market participants also seemed to anticipate these banks' vulnerabilities, as the same banks that suffered more after the SVB failure had also underperformed in the previous year. We  investigate which vulnerabilities were anticipated, and which came as a ``surprise'' following the failure of SVB. It is important to note that all information we use, including the unrealized losses in held-to-maturity securities, was available to the market in January 2023 through public regulatory filings. Our analysis suggests that market participants had already factored in vulnerabilities associated with reliance on uninsured deposits and limited cash holdings. However, damages from implied losses and the impact of bank size were more of a surprise. If depositors do not perceive implied losses as problematic and do not run on the bank, stock investors may have had little reason to worry about them. Although, this changed abruptly with the failure of SVB, which aligns with the information view of banking panics (\textcite{DangGortonHolmstrom2020}), where 
bank runs are caused by abrupt shifts in depositors' risk perceptions.\footnote{According to the information view, a sudden realization of risks in bank assets can be triggered by an arrival of new information such as the failure of a large institution or negative news on aggregate fundamentals (\textcite{Gorton1988}).} 

The SVB episode has similarities with the Savings and Loans (S\&L) crisis and the GFC, as well as important differences. When inflation reached historic levels in the United States, the Federal Reserve started to increase interest rates in October 1979. S\&Ls specialized in residential mortgages and issued long-term fixed-rate loans. This exposed S\&Ls to significant interest rate risk like SVB. However, unlike the 2023 crisis, where rapid depositor runs played a critical role, S\&L crisis was a relatively slow-moving process. Faced by mounting losses, regulators relaxed regulation and exercised forbearance, hoping the S\&Ls could outlast out of their balance sheet problems. This gave S\&Ls the opportunity to gamble for resurrection, resulting in excessive risk taking and imprudent real estate lending. Over time, what started from interest rate risk turned into a crisis of bad loans and credit risk. 

During the GFC, low quality loans, complexity of financial products and the runs in wholesale funding markets played a major role. The new regulatory framework of Basel III has tried to address the fragility of wholesale funding by giving favorable treatment for deposits, which were viewed as a more stable source of funding. The SVB episode showed that uninsured deposits can also be a major source of bank runs. Moreover, even the safest and most liquid securities like Treasuries can also experience significant losses due to interest rate risk. While held-to-maturity accounting can hide these losses on banks' balance sheets, the losses can crystallize when banks facing with urgent liquidity needs are forced to sell the securities at market prices.

Overall, the unrealized losses in securities, the vulnerability of uninsured deposits to runs, and the interaction between the two were new factors during the \ac{SVB} episode. The neglected risk by investors further underscores the challenges for policymakers in crafting robust stress scenarios that comprehensively account for all pertinent vulnerabilities beforehand.  These are important insights for policy makers and will help them in designing new policies to make the financial system more resilient.

\bigskip

\textbf{Related literature:} The paper is related to the vast literature on bank runs (e.g., \textcite{Diamond1983}, \textcite{ChariJagannathan1988}, \textcite{CalomirisKahn1991}, \textcite{AllenGale1998}, and \textcite{goldstein2005demand}) and their origins (\textcite{Gorton1988}, \textcite{CalomirisGorton1991}). 
It also contributes to the literature on the channels of contagion, such as direct exposures via interbank linkages (\textcite{AllenGale2000});
information contagion (\textcite{chen199}, \textcite{AcharyaYorulmazer2008}); and illiquidity and asset prices
(\textcite{DiamondRajan2001}, \textcite{GortonHuang2004}, and \textcite{cifuentes2005}).

The paper is also related to the literature on banks' interest rate risk. A recent study by \textcite{drechsler2021banking}  suggests that banks’ deposit franchise can act as a natural hedge against interest rate risks, while \textcite{DrechslerSavovSchnablWang2023} argue that a bank run equilibrium can emerge when banks heavily rely on uninsured deposits during rate increases. 
Investigating  the 2022 monetary tightening period, \textcite{jiang2023LimitedHedge} find significant exposures to interest rate risk with limited use of hedging (see also \textcite{Begenau2015}).

Several other studies examine the implications of \ac{SVB}'s failure and the subsequent banking crisis. \textcite{metrick2024failure} documents the failure of \ac{SVB} and how its balance sheet contributed to the bank run of 2023. \textcite{jiang2023} and \textcite{FlannerySorescu2023} document substantial mark-to-market losses spread across the banking sector. \textcite{jiang2023} assess how the run by uninsured depositors led to bank insolvency due to the realization of hidden mark-to-market losses. 
\textcite{HaddadHartmanMuir2023} present a theoretical model that characterizes the role of rapid rate hikes in bringing this vulnerability and offer  empirical evidence supporting their predictions.\footnote{Focusing on the spillover mechanism, \textcite{BenmelechYangZator2023} analyze the role of bank branch presence and \textcite{cookson2023} assess the impact of social media exposures. \textcite{DercoleWagner2023} examine the stock price reactions of environmentally responsible stocks to the failure of SVB.} 
\textcite{caglio2023} present findings on depositors' flight to safety, whereby funds shift away from regional banks towards larger banks.

In contrast to papers that focus on the effect of a specific factor on contagion, our study provides a comprehensive assessment of multiple factors contributing to negative spillovers. We emphasize the unique characteristics of the 2023 crisis compared to the \ac{GFC}, offering valuable insights for regulatory objectives.  Additionally, our analysis distinguishes between the immediate and medium-term effects, uncovering the emergence of the too-big-to-fail  effect as concerns spread across the entire banking system. Finally, we find that investors partially anticipated vulnerabilities associated with uninsured deposits, but other factors, such as those related to unrealized losses and bank size, appeared to come as a surprise.

\section{Background and Conceptual Framework}
\label{sec:background}
First, we provide a background and a summary of the events that led to the failure of \ac{SVB}. Then, we discuss the factors that might have played a role in the failure of \ac{SVB} such as interest rate risk and implied losses, liquidity risk and unsecured deposits, and the rollback of regulation.

\subsection{A summary of the SVB failure}
\ac{SVB} was founded in 1983 with a focus on the needs of startup companies. Its main strategy was to collect deposits from businesses financed through venture capital while it expanded into banking and financing venture capitalists. During the 1980s, \ac{SVB} grew with the local high-tech economy, which persisted  during the 1990s as startups during the dot-com bubble provided an influx of business for the bank.  In the 2000s, \ac{SVB} entered into the private banking business building on earlier experience and relationships with wealthy venture capitalists and entrepreneurs, and enjoyed an international expansion which continued in the 2010s. In 2015, the bank stated that it served 65\% of all U.S. startups. 

During the loose monetary policy of the pandemic-era and the investment boom in private technology companies, \ac{SVB} enjoyed a significant increase in its deposits and asset size, where, just in 2021, deposits surged from \$102 billion to \$189 billion.\footnote{Deposits and assets of \ac{SVB} tripled between 2019 and 2021.} Awashed with deposits, \ac{SVB} increased its investment in highly-rated government bonds in its portfolio to \$120 billion of which \$91 billion consisted of fixed-rate mortgage bonds with very long maturities. This, in turn, exposed \ac{SVB} to interest rate risk. On the liability side, regulatory filings estimated that more than 90\% of its deposits were uninsured, which exposed the bank to funding liquidity risk and a depositor run. 

In response to the spike in inflation following the COVID-19 pandemic, the Federal Reserve began to increase interest rates in March 2022. This led to \ac{SVB} incurring heavy losses on its bond portfolio. While \ac{SVB} deposits dropped for four straight quarters prior to its failure, deposit withdrawal was faster than expected in February and March of 2023. As a result, \ac{SVB} had to liquidate some of its HTM assets, and incurred a loss of \$1.8 billion. To cover these losses, \ac{SVB} announced a capital raise of \$2.25 billion in the week of its failure. On March 8th Moody's downgraded SVB\footnote{\textcite{BootMilbournSchmeits2006} analyze the coordination role of credit ratings.} and it was apparent on March 9th that \ac{SVB} was not being able to raise the needed capital. This exacerbated a depositor run with withdrawals of \$42 billion, equal to a quarter of bank deposits, on a single day. On March 10th, \ac{SVB} failed as a result of the bank run and was put into receivership by the FDIC. On March 12th, in a joint statement by the Secretary of the Treasury, Federal Reserve Chairman and the FDIC Chairman, a blanket guarantee for insured and uninsured deposits at \ac{SVB} was introduced.

\subsection{Factors}
In this section, we discuss factors which had a role in the failure of \ac{SVB} and the subsequent spillover effects such as interest rate risk and implied losses, liquidity risk arising from uninsured deposits, and an erosion of the regulatory framework governing mid-sized banks.

\subsubsection{Interest rate risk and implied losses}

To combat inflation, the Federal Reserve began hiking interest rates in March 2022. \ac{SVB} had significant exposure to interest rate risk in its bond portfolio. \ac{SVB}'s bond portfolio was separated into HTM and AFS securities. HTM securities are planned to be held until maturity and can be carried at their nominal par value since they are being held until they are repaid in full. On the other hand, AFS securities are marked to market. At the end of 2022, \ac{SVB} had \$91.3 billion in HTM securities with very long maturities and \$26.1 billion in AFS securities. Increasing interest rates led to losses in the market value of its bonds with significant unrealized losses for its HTM securities. The rise in interest rates and the resulting losses in the securities portfolio of SVB, some of which were hidden due to HTM accounting, was a major contributor in the demise of SVB \parencite{jiang2023}.

\subsubsection{Liquidity risk due to unsecured deposits}

During the \ac{GFC}, a key lesson learned was the importance of differentiating between wholesale funding and deposits when assessing funding risk.  While wholesale funding was highly vulnerable to runs during the crisis, deposits turned out to be relatively ``sticky'' and less likely to be withdrawn by panicked investors (see, e.g., \textcite{ivashina2010bank} and \textcite{hanson2015banks}).  This insight has been incorporated into regulatory reforms aimed at improving the resilience of the banking system such as the \ac{LCR} requirement of Basel III (\textcite{BIS13}). In particular, the \ac{LCR} requirement mandates that banks must have sufficient high quality liquid assets (HQLA) to meet net cash outflows for 30 days under a stress scenario.\footnote{HQLAs mainly consist of reserve balances, Treasuries, agency debt and agency MBS.} \ac{SVB}'s bond portfolio, which constituted a major part of its assets, would be considered as HQLA in the context of \ac{LCR}.\footnote{Due to the rolling back of regulation, as we discuss next, SVB was not subject to the \ac{LCR} requirement. However, \textcite{Nelson2023} estimates that SVB would have satisfied the \ac{LCR} requirement since a majority of its asset portfolio would be considered to be highly liquid.} 
Nevertheless, the securities in its portfolio had interest rate risk exposure. Even though both cash and most of the securities were treated to be highly liquid in the \ac{LCR} framework, cash ended up being significantly more valuable to banks, as we discuss in Section \ref{sec:whatwasdifferent}.

\ac{LCR} assumes a lower outflow rate for deposits compared to market funding, reflecting the greater stability and reliability of deposit funding. \ac{SVB} largely financed itself via uninsured deposits, which turned out to be subject to severe runs even though they were expected to be less ``flighty'' compared to other wholesale funding sources under \ac{LCR}.  Heavy reliance on uninsured deposits exposed \ac{SVB} to funding liquidity risk. As we argued above, potential losses becoming actual losses due to deposit withdrawals played a crucial role in the demise of \ac{SVB}.

\subsubsection{Rolling back regulation for mid-sized banks}

After the \ac{GFC}, the Dodd-Frank Act introduced major reforms to the bank regulatory framework. However, on May 24th, 2018, The Economic Growth, Regulatory Relief, and Consumer Protection Act was signed into law and eased the regulations for mid-sized banks by raising the threshold for systemically important financial institutions from \$50 billion to \$250 billion.\footnote{The act gave the Federal Reserve discretion in determining regulations for financial institutions with assets between \$100 billion and \$250 billion.} One consequence of the regulatory rollback was that \ac{SVB} was not subject to the \ac{LCR} requirement. Furthermore, despite interest rate risk being the major factor in the failure of SVB, regulatory stress tests did not incorporate an interest rate risk scenario.\footnote{See \url{https://www.federalreserve.gov/publications/2023-Stress-Test-Scenarios.htm}.}

\section{Data and Methods}

\subsection{Data}
\label{sec:data}
We start with the Federal Reserve Bank of New York's permanent company number (PERMCO)-RSSDID crosswalk.\footnote{Available at \url{https://www.newyorkfed.org/research/banking_research/crsp-frb}.} This crosswalk provides the PERMCO identifier for most of the regulated banks in the United States. Using the Center for Research in Security Prices (CRSP) database, we link the PERMCO to each firm's stock ticker as of 2022q4. The linked list dataset of RSSDID and tickers has 327 banks.

Next, we construct a dataset using these tickers of daily close stock price data from Yahoo! Finance for February 1, 2022 until May 25, 2023. We are able to pull this data for 324 of the tickers. We construct daily returns based on these measures, and  a measure of market returns based on close price of the S\&P 500. We also construct a daily return measure for banks based on the Dow Jones U.S. Banks Index (DJUSBK). 

Then, for each of the 327 RSSDIDs in our initial list, we identify all subsidiaries of the bank using the National Information Center (NIC)'s relationship file.\footnote{Available at \url{https://www.ffiec.gov/npw/FinancialReport/DataDownload}.} For each subsidiary, if they are regulated by the FDIC, we identify their FDIC Certificate Number, and collect their total deposits, total uninsured deposits, and total assets using the FDIC's BankSuite tool.\footnote{Available at \url{https://banks.data.fdic.gov/bankfind-suite}} This is a sample of 318 top holders and banks.\footnote{We omit Silvergate Bank from this analysis. This is due to its unique features, including its relatively small size, single industry concentration in crypto-currencies, and heavy reliance on FTX as its primary depositor.}

Finally, for those top holder companies who report FR Y-9C data and for commercial banks that report Call Report data, we construct several financial balance sheet measures from their 2022q4 report. These include total assets, cash, securities, HTM and AFS securities, mark-to-market losses on HTM securities based on the difference between their book values and fair values, Tier 1 capital, and non-performing loans. This is a sample of 224 top holders and banks.\footnote{A number of the original 318 banks report FR Y-9SP data because they have less than \$3 billion in assets. The FR Y-9SP reporting form reports significantly less than the FR Y-9C form for our main sample. As a result, we currently exclude the banks that only report FR Y-9SP data due to data limitations.} Our final sample is described in Table \ref{tab:summary_stats}. For several of our banks, neither their regulatory Tier 1 capital ratios nor amounts of Tier 1 capital are reported. In regressions using these measures, we lose 2 and 8 observations, respectively.\footnote{This crosswalked data (and replication code for the paper) is available at \url{http://github.com/paulgp/bank_returns/}.}

\begin{table}[th]
    \centering
    \caption{Sample Composition and Summary Statistics. This table reports summary statistics of the main bank characteristics in our sample, measured as of 2022q4. }
    \label{tab:summary_stats}
   \footnotesize
\makebox[\linewidth][c]{\begin{tabular}{lrrrr}
\toprule
Variable & Observations & Mean & Median & SD \\ 
\midrule
Assets (000) & 224 & 101,688,402 & 9,559,211 & 405,970,115 \\ 
Cash / Assets & 224 & 9.2\% & 8.1\% & 7\% \\ 
Securities / Assets & 224 & 20\% & 18\% & 11\% \\ 
Liquid Assets / Assets & 224 & 29\% & 27\% & 13\% \\ 
Unrealized HTM Losses / Tier 1 Capital & 222 & 0.21\% & 0\% & 2.2\% \\ 
held-to-Maturity Securities / Assets & 224 & 3.2\% & 0.33\% & 5.9\% \\ 
Uninsured Deposit Share & 224 & 39\% & 38\% & 18\% \\ 
Tier 1 Capital Ratio & 216 & 13\% & 12\% & 2.9\%\\ 
\bottomrule
\end{tabular}}
\end{table}

\subsection{Cumulative Returns in Excess of S\&P 500}
We construct daily returns $r_{it}$ for each bank $i$ in period $t$ using the closing price for each day. For our baseline analysis assessing the short-term impact, we  calculate  bank $i$'s cumulative return starting from February 1, 2023:
\begin{equation}
    R_{i}^{early} = \prod_{t \geq 2023-02-01}(1+r_{it}) - 1  \label{eq:CumReturn}
\end{equation}
with March 17, 2023, a week after the \ac{SVB} bank run, as the final date and is referred to as  ``early returns.'' 
To assess the medium-term impact, we similarly define ``late returns,'' denoted as $R_{i}^{late}$, with May 25, 2023 as the final date.\footnote{In general, we refer to any cumulative returns starting from February 1, 2023, as ``2023 returns.''}

We then consider the excess return for this cumulative return above the cumulative return of the S\&P 500 over the same time period\footnote{Our results remain similar when using $\beta$-adjusted excess returns, following \textcite{campbell1998econometrics}. See the Appendix Tables for regression results using the $\beta$-adjusted returns.} (denoted as $R^{early}_{m}$ and $R^{late}_{m}$, respectively), that is,
\begin{equation}
    \tilde{R}^{early}_{i} = R^{early}_{i} -R^{early}_{m}. \label{eq:ExcessReturn}
\end{equation}

This measure can be interpreted as the cumulative return difference between investing a dollar long in bank $i$'s stock and a dollar short in the S\&P500 on February 1, and holding the position until the final date. This differs slightly from traditional event study approaches by focusing on excess geometric returns, rather than arithmetic excess returns. We also construct ``2022 returns'' between February 1, 2022 to January 31, 2023 to capture the pre-panic returns.  

We consider how the excess return $\tilde{R}_{i}$ covaries with various bank-level characteristics using linear regression. These regressions are run in the cross-section, using robust standard errors.

\section{Results}

We first present the overall trends in bank stock returns since the commencement of the rate hikes by the Federal Reserve in early 2022. Next, we analyze the vulnerability factors associated with spillover effects following the \ac{SVB} failure. Finally, we discuss whether market participants had anticipated such vulnerabilities ex ante. 

\subsection{Overall trends}

\begin{figure}
    \centering
    \includegraphics[width=\linewidth]{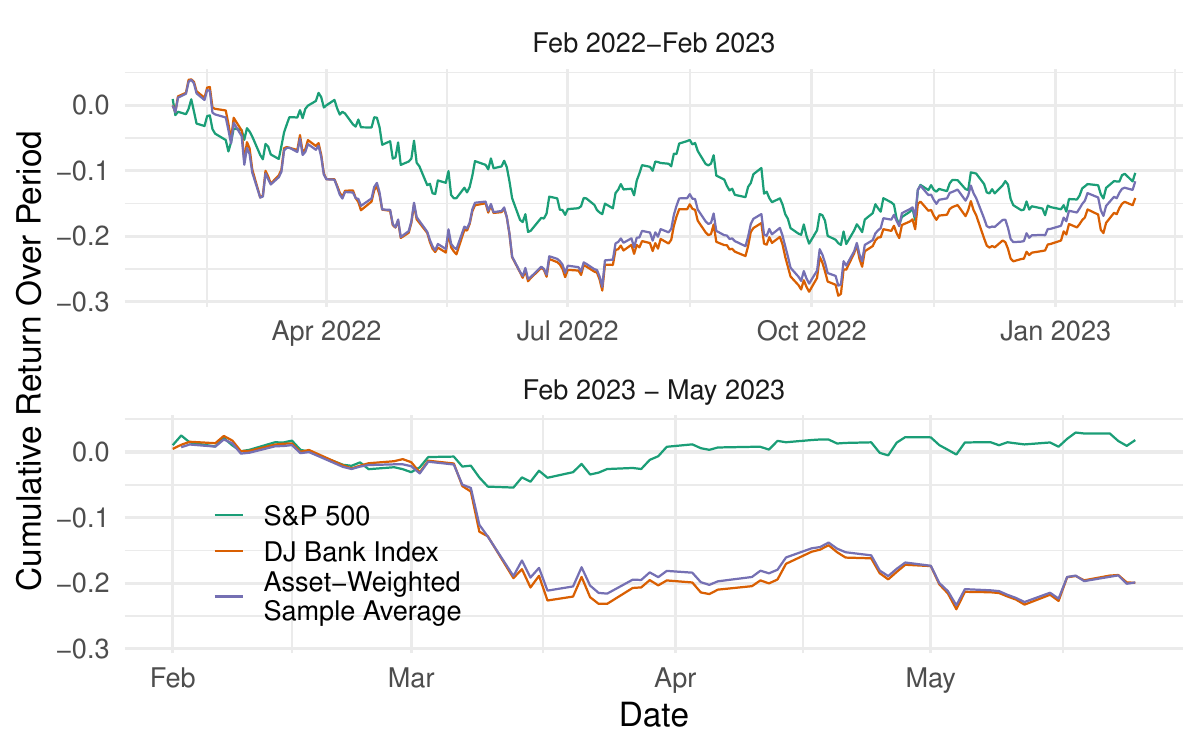}
    \caption{Aggregate cumulative returns for the S\&P 500 and banking sector. This figure plots cumulative returns for three sets of stock indices: the S\&P 500 index, the Dow Jones U.S. Banks Index, and the asset-weighted average of the banks in our sample. The top panels reports the cumulative return starting in February 1, 2022, until January 31, 2023. The bottom panel reports the cumulative return from February 1, 2023, until May 25, 2023. }
    \label{fig:enter-label}
\end{figure}

Figure \ref{fig:enter-label} displays the cumulative returns for the S\&P 500 index and the banking sector. The top panel plots the 2022 returns to illustrate the trends observed over the 12 months preceding the \ac{SVB}'s failure, while the bottom panel shows the 2023 returns focusing on the subsequent 3 months after the event. We plot two measures of banking sector return: the Dow-Jones U.S. Bank Index (DJUSBK) and the asset-weighted average of our sample banks. The two sectoral returns exhibit a correlation of 0.99 in 2022.

Since the start of the rate hikes on March 17, 2022, there was a general decline in stock prices. Notably, the banking sector initially exhibited even weaker performance compared to the rest of the market, as indicated in the top panel. However, this trend partially reversed later in the same year as the stock market started to recover. By early 2023, the cumulative impact of rate hikes was similar for banks and the overall market. 

\begin{figure}[th]
    \centering
    \includegraphics[width=\linewidth]{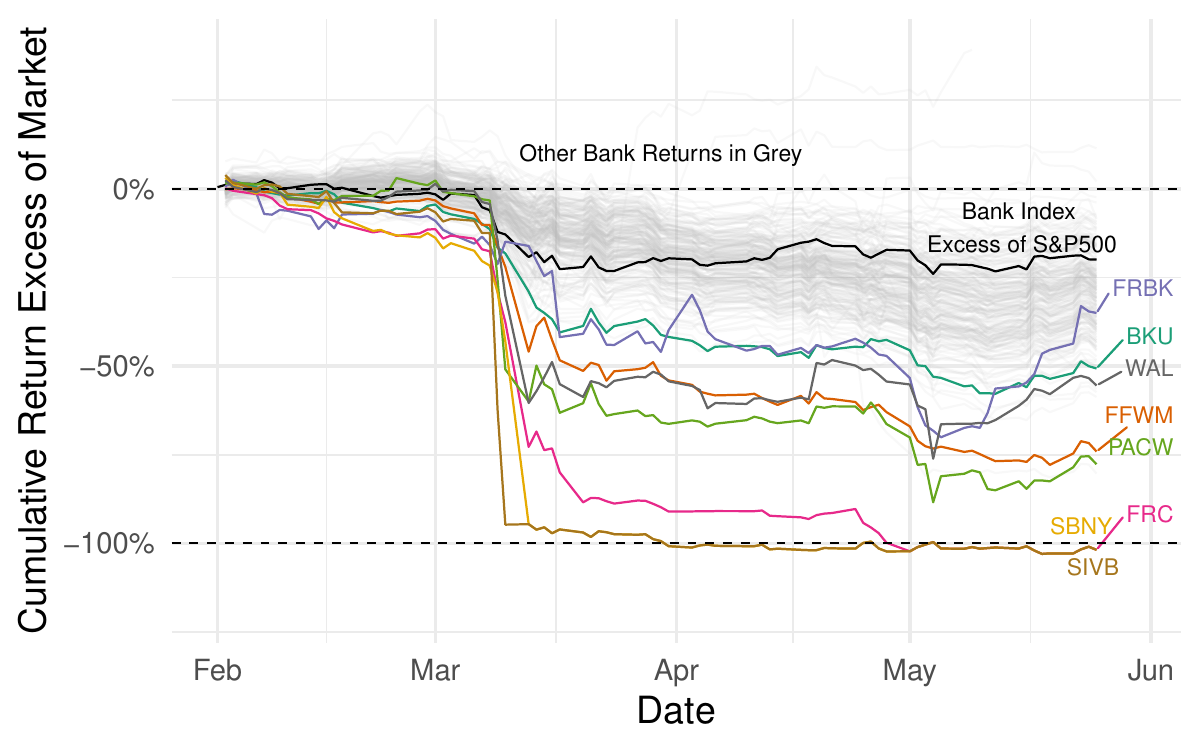}
    \caption{Bank Returns Over Time for Select Banks. This figure plots cumulative returns  in excess of the S\&P 500 index from February 1, 2023, until May 25, 2023. We highlight the 8 banks with the largest cumulative declines as of March 17th and the Dow Jones U.S. Bank Index's abnormal returns. The rest of the banks in our sample are plotted in grey. Stocks are denoted by their ticker symbol; Silicon Valley Bank's ticker is SIVB. }
    \label{fig:over_time}
\end{figure}

Following the bank run of \ac{SVB} in March 2023, the relative performance of the banking sector diverged sharply from the rest of the market, as depicted in the bottom panel of Figure \ref{fig:enter-label}. While the S\&P index did not show distinct signs of stress following the \ac{SVB} failure, the banking sector as a whole declined by 20\%. Furthermore, in Figure \ref{fig:over_time}, which plots individual banks' 2023 returns in the same time period as the Panel B of Figure \ref{fig:enter-label}, we observe a noticeable increase in cross-sectional variation in banks' stock performance after the \ac{SVB} bank run. The distribution of bank stock returns remained relatively tight prior to the FDIC takeover of \ac{SVB} but widened significantly afterwards, indicating the presence of heterogeneous spillover effects. We highlight the eight banks with the largest declines as of March 17th. Some of these banks, including SVB and Signature Bank (SBNY), did not post market prices following their FDIC takeovers until later, and hence had fixed prices for several days. Several other banks, notably First Republic (FRC), also experienced substantial immediate declines.  Our empirical analysis primarily aims to identify the factors contributing to these differential spillovers.

\begin{figure}
    \centering
    \hfill
       \begin{subfigure}[b]{0.45\linewidth}
         \centering
         \includegraphics[width=\linewidth]{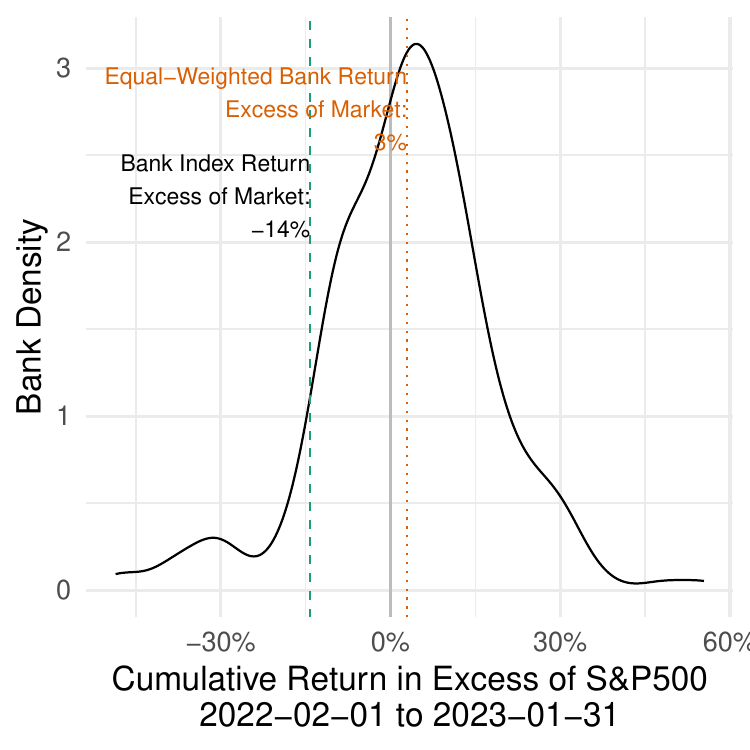}
         \caption{CAR since 2022-02-01}
         \label{fig:early}
     \end{subfigure}
    \hfill
    \begin{subfigure}[b]{0.45\textwidth}
         \centering
         \includegraphics[width=\textwidth]{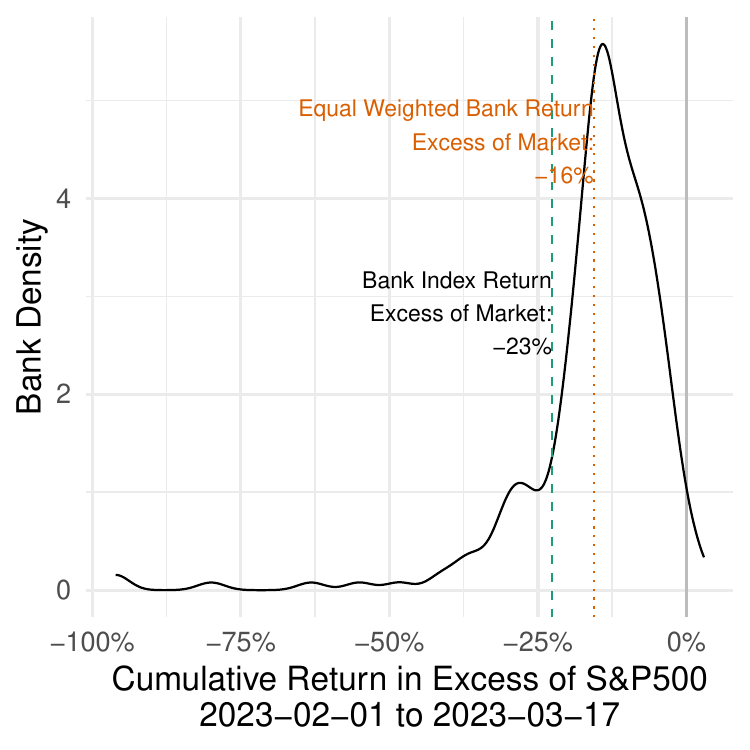}
         \caption{CAR since 2023-02-01}
         \label{fig:SVB}
     \end{subfigure}

            \begin{subfigure}[b]{0.45\textwidth}
         \centering
         \includegraphics[width=\textwidth]{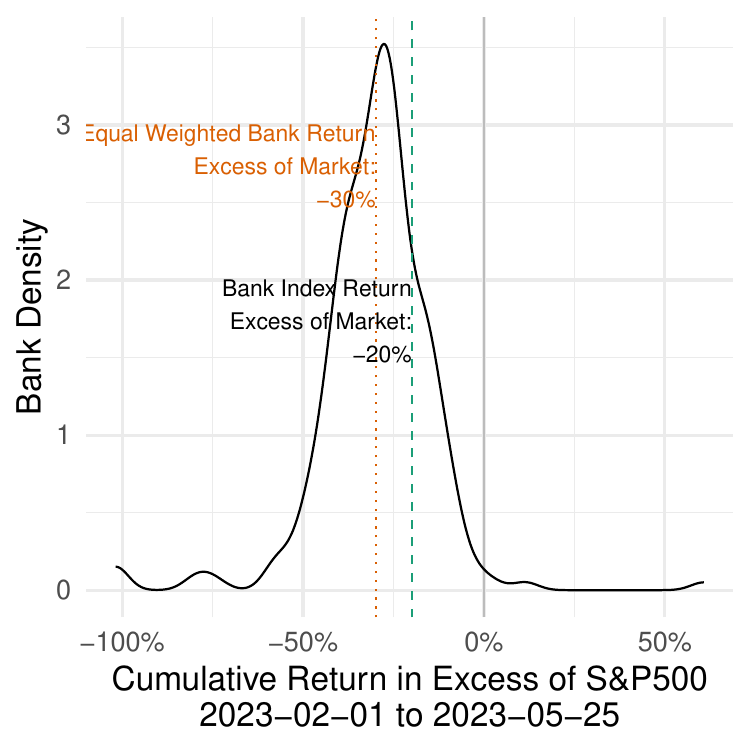}
         \caption{CAR since 2023-02-01}
         \label{fig:long}
     \end{subfigure}
         \begin{subfigure}[b]{0.45\textwidth}
         \centering
         \includegraphics[width=\textwidth]{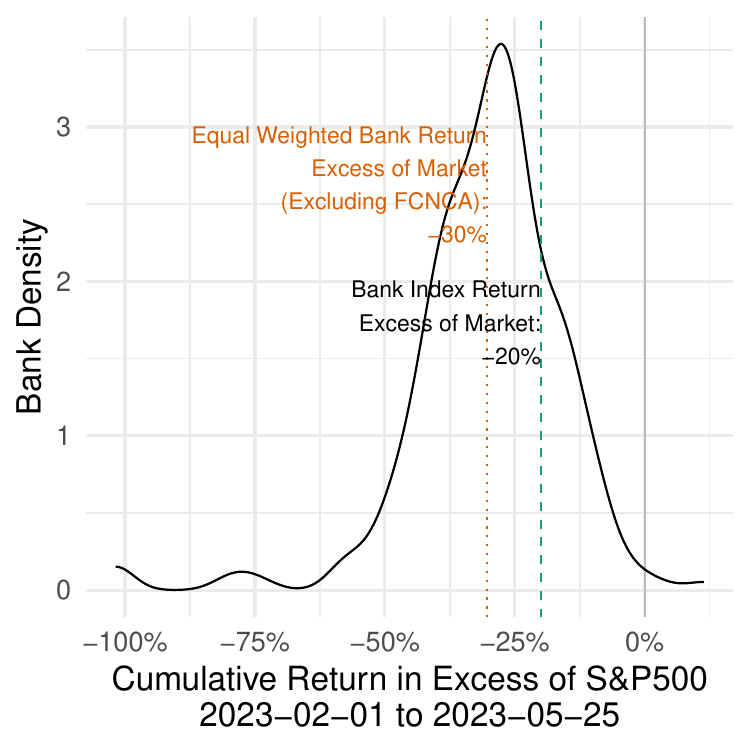}
         \caption{CAR since 2023-02-01\\ Excluding FCNCA}
         \label{fig:long2}
     \end{subfigure}
    \caption{Distribution of abnormal returns by time period. This figure plots the distribution of cumulative returns in excess of the S\&P 500 index for the banks in our sample over three periods. In Panel (a), we report the distribution of cumulative returns as of January 31, 2023, measured since February 1, 2022.  In Panel (b), we report the distribution of cumulative returns as of March 17, 2023, measured since February 1, 2023. In Panel (c), we report the distribution of cumulative returns as of May 25, 2023, measured since February 1, 2023. Panel (d) repeats Panel (c) excluding First Citizens Bankgroup. We additionally report the Dow Jones U.S. Bank Index cumulative return for each period as a vertical line, and the equal weighted mean of the banks in our sample for each period.}
    \label{fig:agg_dist_returns_period}
\end{figure}

Figure \ref{fig:agg_dist_returns_period} plots the full distribution of bank stock returns assessed across different time periods. Panel \ref{fig:early}, based on cumulative returns from February 1, 2022 to January 30, 2023, indicates that banks' relative returns compared to the market index were symmetrically distributed around 0. This suggests the absence of a discernible aggregate shock unique to the banking industry. However, following the bank run, the distribution of bank excess returns shifted towards the left and exhibited wider dispersion at the left tail. Panel \ref{fig:SVB} presents the distribution of ``early'' returns from February 1, 2023, to March 17, 2023 (a week after the \ac{SVB} bank run), aiming to assess the immediate spillover effect. Panel \ref{fig:long} displays the ``late'' returns up to May 25, capturing the medium-term effect, which shows a further shift to the left from that in Panel \ref{fig:SVB}, with the average return declining from -16\% to -30\%.

Notably, in Panel \ref{fig:long2}, we plot the same figure as \ref{fig:long} but exclude an outlier bank with a significant positive return.\footnote{First Citizens Bank experienced a significant positive return following its acquisition of \ac{SVB}'s assets and deposits, see Figure \ref{fig:returns_v_assets_late}.} The shapes of the early and late return distributions in Panels \ref{fig:SVB} and \ref{fig:long2} appear similar. However, the composition of these two distributions differs significantly. As indicated by the difference between the value-weighted return (in blue) and the equal-weighted return (in red), banks with large assets generally underperformed compared to small banks in 2022 (Panel \ref{fig:early}) and immediately after the \ac{SVB} bank run (Panel \ref{fig:SVB}). Interestingly, this trend reversed in Panel \ref{fig:long}, with the value-weighted returns significantly higher than the equal-weighted returns, indicating large banks outperforming small banks. We delve into the implications of these effects in Section \ref{subsec:size}.

\begin{figure}
    \centering
    \includegraphics[width=0.8\linewidth]{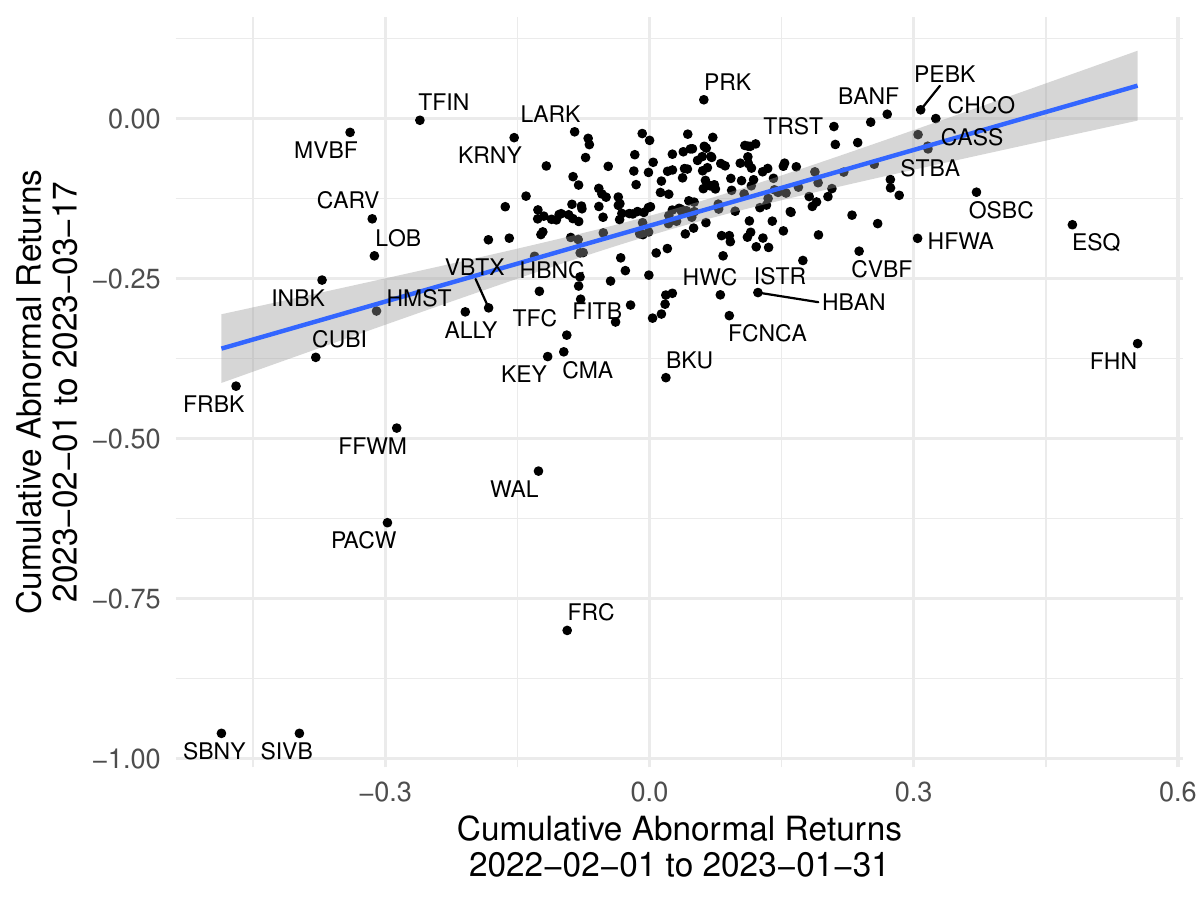}
    \caption{2022 bank excess returns vs 2023 bank excess returns. This figure compares cumulative returns in excess of the S\&P 500 index for the banks in our sample for the 2022 period versus the early 2023 period. On the x-axis, we plot the cumulative returns in excess of the S\&P 500 index for the banks in our sample from February 1, 2022 to January 31, 2023. On the y-axis, we plot the cumulative returns in excess of the S\&P 500 index for the banks in our sample from February 1, 2023 to March 17, 2023. We also plot the least squares line of best fit. Stocks are denoted by their ticker symbol; Silicon Valley Bank's ticker is SIVB.}
    \label{fig:2022_v_2023}
\end{figure}

Lastly, we examine whether banks experiencing larger negative spillovers also demonstrated poorer performance in 2022. To do so, we compare the cumulative returns before and immediately after the \ac{SVB} bank run, i.e., 2022 returns and 2023 early returns, as in Figure \ref{fig:2022_v_2023}. Notably, there is a robust correlation between these two sets of returns across banks, with a bivariate $R^{2}$ of 0.18, and a regression coefficient of 0.31 with a corresponding $t$-statistic of 4.5. As the latter period captures the immediate negative spillovers associated with the \ac{SVB} failure, this pattern suggests that market participants, to some extent, anticipated and factored in the risks associated with the banking crisis of 2023 ahead of time.

\subsection{Spillover effects -- driving factors }

We now examine the specific factors that explain the heterogeneous magnitudes of price declines resulting from the SVB failure. Models of bank runs suggest that a bank's vulnerability is linked to various factors including asset quality, asset liquidity, funding outflow risk, and capitalization, among others.\footnote{See models such as \textcite{Diamond1983,AllenGale1998,EisenbachKeisterMcAndrewsYorulmazer2014,rochet2004coordination,goldstein2005demand}.} 

We begin by discussing two vulnerabilities: uninsured deposits and unrealized losses on securities. Then, we explore whether other conventional factors such as asset liquidity, leverage, asset quality or size are associated with the spillovers observed in this particular event. Finally, we discuss which of these vulnerabilities investors may have anticipated, and which factors emerged as a ``surprise'' following the occurrence of the bank run.

\subsubsection{Unique factors - uninsured deposits and unrealized losses on securities}
As discussed in Section \ref{sec:background}, the primary reason for the failure of \ac{SVB} was concerns over unrealized losses in its security holdings, which prompted uninsured depositors to withdraw their funds. In normal circumstances, banks can use their liquid securities as buffers to alleviate shocks from funding outflows. However, with unrealized losses due to rate increases, liquidation of the HTM securities, while still ``liquid'', would force banks to mark losses, exacerbating financial distress.

Deposits, even uninsured ones, typically serve as a reliable source of funding for banks as they are often considered less flighty than other market funding sources (\textcite{ivashina2010bank,hanson2015banks,bai2018measuring}). Thanks to the deposit franchise, banks can also attract deposits at a lower cost than other sources (\textcite{BERGER1987501, gale2020bank,gatev2006banks, drechsler2017deposits, drechsler2021banking}). This conventional wisdom is also reflected in the liquidity regulation, where the \ac{LCR} adopts relatively low run-off rates for deposit sources than other market funding (\textcite{BIS13}). 

However, during the 2023 crisis, banks could not easily exploit the deposit franchise as they needed to offer significant interest rates to attract or retain depositors (\textcite{KangLuckPlosser2023}). 
They also faced competition from other investment options, such as money market funds, which had access to the Fed's reverse repo facility and were perceived as safe while offering high interest rates. 

\begin{figure}[th]
    \centering
    \includegraphics[width=0.8\linewidth]{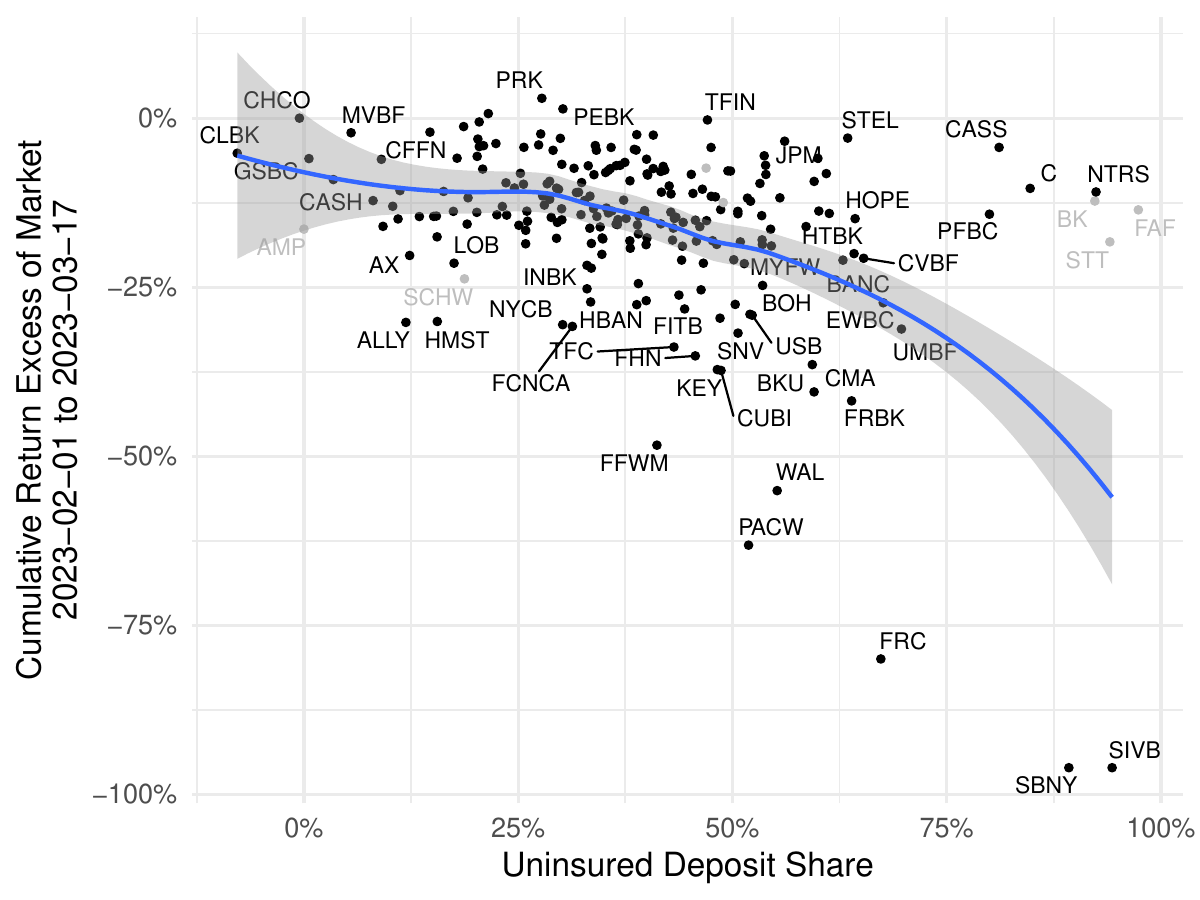}
    \caption{Cumulative returns vs. uninsured deposit share. This figure plots the cumulative returns in excess of the S\&P 500 index for the banks in our sample from February 1, 2023 to March 17, 2023 against the uninsured deposit share (uninsured deposits as a share of total deposits) measured in 2022q4. See Section \ref{sec:data} for details on the data construction. The fitted line is a local linear regression (loess) curve. Stocks are denoted by their ticker symbol; Silicon Valley Bank's ticker is SIVB. }
    \label{fig:returns_v_uninsured_deposit_share}
\end{figure}

We begin by examining the link between banks' reliance on uninsured deposits and negative stock returns in the aftermath of the \ac{SVB} failure. As shown in  \Cref{fig:returns_v_uninsured_deposit_share}, which compares banks' use of uninsured deposits to their stock performances (i.e., early returns), a clear negative relationship between the two can be observed.\footnote{In assessing this relationship in  \Cref{fig:returns_v_uninsured_deposit_share}, we grey out and omit BHCs with atypical funding structures. These include custodian banks (Bank of New York Mellon Corp (BK) and State Street Corp (STT)) and 
insurance providers (First American Financial Corp (FAF)).} Notably, the three eventually failed banks — \ac{SVB}, SBNY, and FRC — had exceptionally high uninsured deposit shares, with shares of 94\%, 89\%, and 67\%, respectively, compared to the average share of 39\%.

\begin{figure}[ht]
    \centering
     \begin{subfigure}[b]{0.49\linewidth}
         \centering
         \includegraphics[width=\linewidth]{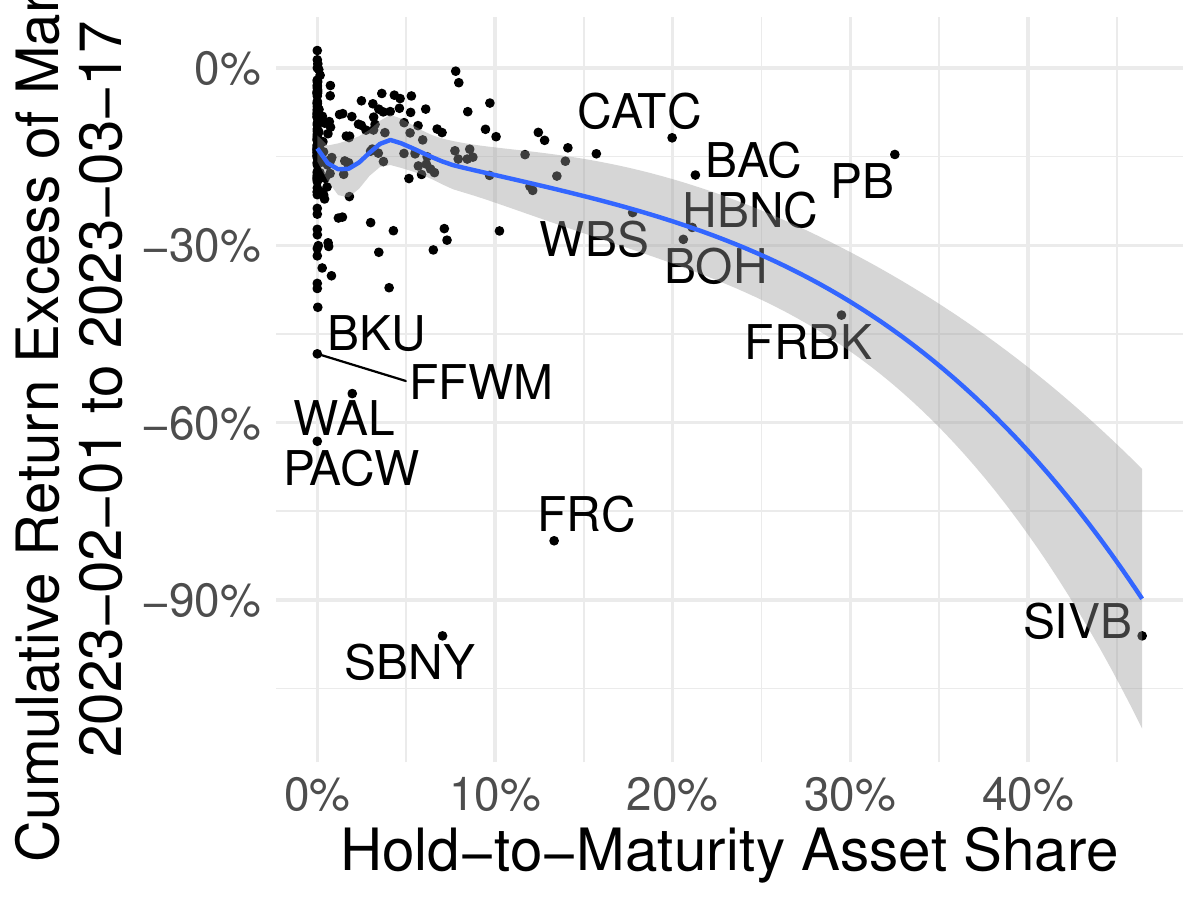}
         \caption{Held-to-maturity asset share }
     \end{subfigure}
     \begin{subfigure}[b]{0.49\linewidth}
         \centering
         \includegraphics[width=\linewidth]{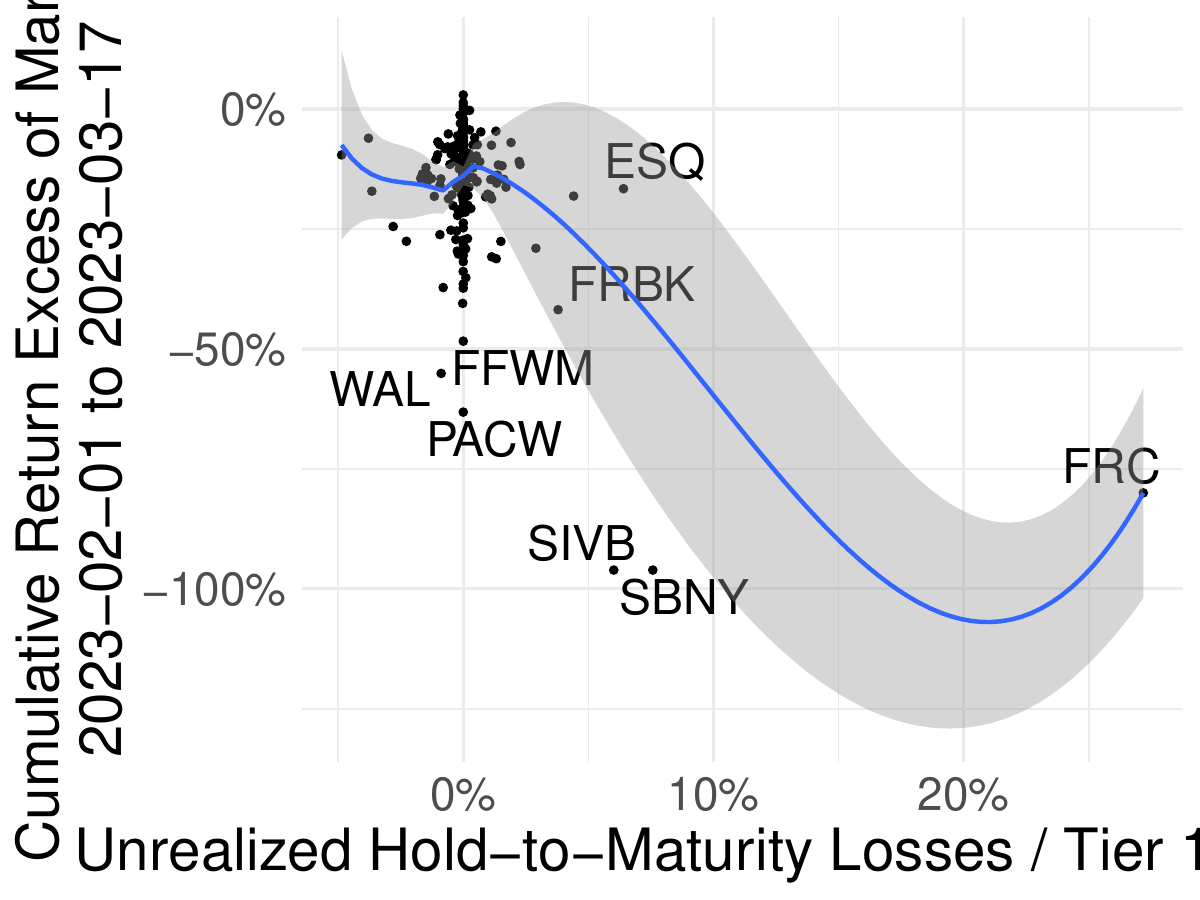} 
         \caption{Unrealized HTM losses / Tier 1 Capital }
     \end{subfigure}
    \caption{Early cumulative returns explained by held-to-maturity assets. Panel (a) plots the cumulative returns in excess of the S\&P 500 index for the banks in our sample from February 1, 2023 to March 17, 2023 against the held-to-maturity asset share (held-to-maturity assets as a share of total assets) measured in 2022q4. Panel (b) plots against  unrealized held-to-maturity losses scaled by Tier 1 capital. See Section \ref{sec:data} for details on the data construction.  The fitted lines are local linear regression (loess) curves. Stocks are denoted by their ticker symbol; Silicon Valley Bank's ticker is SIVB. }
    \label{fig:returns_v_htm_share}
\end{figure}

Next, we investigate the relationship between unrealized losses for HTM securities and negative stock returns. This relationship is plotted in \Cref{fig:returns_v_htm_share}, which includes two panels. Panel A shows the ratio of a bank's HTM securities to its total assets, while Panel B displays the ratio of a bank's unrealized losses in HTM securities to its Tier 1 capital.

In Panel A, it can be seen that many banks have very small amounts of HTM securities, resulting in the HTM/asset shares being clustered around 0. However, there are several outliers with a large HTM/asset share, the most notable being \ac{SVB}. Conditioning on larger shares of HTM securities, we observe a negative association between the shares of HTM securities and the stock returns.

In Panel B, unrealized HTM losses to Tier 1 capital are predominantly clustered around 0. This is due to many banks having minimal exposure to HTM securities, thereby constraining their potential losses.  Additionally, certain banks may have opted for securities with lower duration risk compared to FRC or \ac{SVB}. However, there is a negative association between unrealized losses and the stock returns as illustrated in the figure. It is worth noting that the implied losses from HTM securities, once recognized, would have reduced FRC's Tier 1 capital by almost 30 percent.

\begin{figure}[thbp]
    \centering
    \includegraphics[width=\linewidth]{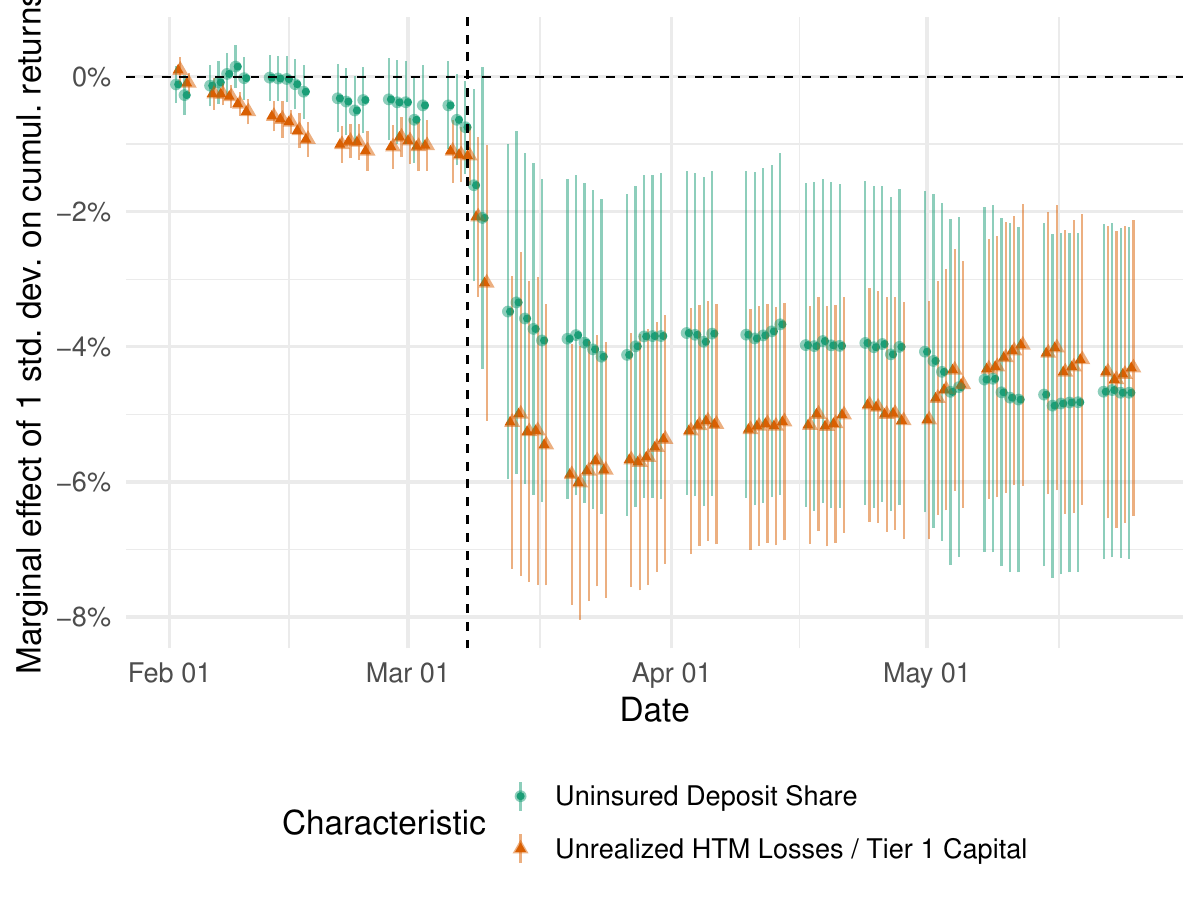}
    \caption{Cumulative returns correlated with uninsured deposits and HTM losses, over time. This graph plots the coefficients for uninsured deposit share and unrealized HTM losses scaled by Tier 1 capital (Column 3 of \Cref{tab:dep_htm}) for every trading day. We omit day fixed effects, which correspond to the constant for each cross-sectional regression. }
    \label{fig:returns_coef_over_time}
\end{figure}

Figure \ref{fig:returns_coef_over_time} illustrates the dynamic association between the two vulnerability factors and banks' stock returns over time. For each date, we regress bank $i$'s cumulative return up to that date (derived from Equation  (\ref{eq:CumReturn})) on a constant term and two vulnerability factors, banks' uninsured deposit share and unrealized HTM losses scaled by Tier 1 capital. These factors are standardized to be mean zero and standard deviation of one. We then plot the estimated coefficients of the respective vulnerability factor for each date. This allows us to assess how the impact of these factors jointly evolved over time. 

The two trends indicate that while bank stock returns were not significantly associated with their reliance on uninsured deposits until early March, investors were aware of the existence of implied losses, which was public information, with declines correlated with them during the beginning of February. However, following the SVB bank run, there arose large and immediate negative spillovers associated with these vulnerability factors, and these effects persisted thereafter, although the magnitude of the correlation declined for unrealized HTM losses.

To quantify the spillover effects following the bank run, we next regress banks' excess early returns (from Equation (\ref{eq:ExcessReturn})) on these bank characteristics.\footnote{Regression results based on late returns are similar to those using early returns and we report them in the Appendix. The only exception is when we assess the impact of bank size, which we discuss in Section \ref{subsec:size}.} To ensure comparability, we standardize the control variables, which allows us to interpret the estimates as the changes in stock returns in response to a 1 standard deviation change in specific vulnerability factors. The regression results are reported in Table \ref{tab:dep_htm}.

Columns 1-3 of Table \ref{tab:dep_htm} represent linear regressions of excess cumulative returns on Uninsured Deposit Share, HTM Asset Share, and Unrealized HTM Losses/Tier 1 Capital, respectively. Notably, all estimates are statistically significant in explaining the negative excess returns. However, the estimate in column 3 for unrealized HTM losses is both statistically and economically the most significant.

In column 4, we include all three variables simultaneously. The results show that while the uninsured deposit share and unrealized HTM losses continue to predict negative excess returns, the actual HTM holdings do not. This finding aligns with earlier intuition, as investors would likely have greater concerns regarding the actual unrealized losses on HTM security portfolios rather than the holdings themselves. The estimates indicate that a one standard-deviation increase in unrealized HTM losses to Tier 1 capital (or uninsured deposit share) would lead to a 4.8 (or 3.5) percentage point lower excess stock returns.\footnote{We present similar results in the Appendix Tables when analyzing $\beta$-adjusted excess returns.}

Note that the presence of unrealized losses in banks' HTM  securities alone may not have a significant impact unless banks are compelled to sell these securities as a result of deposit outflows (\textcite{jiang2023}). 
 This prediction is supported in column 5, where we include the interaction term   between uninsured deposit shares and unrealized HTM losses. The estimate for the interaction term is negative and significant, while unrealized HTM losses by themselves are not significant.

\begin{table}[htbp]
   \caption{\label{tab:dep_htm} Cumulative returns correlated with uninsured deposits and HTM securities. This table reports estimated coefficients of regressions with the cumulative returns in excess of the S\&P 500 index for the banks in our sample from February 1, 2023 to March 17, 2023 as the outcome.  In Column (1), we report the bivariate relationship with the uninsured deposit share (uninsured deposts as a share of total
   deposits) measured in 2022q4. Column (2) reports the coefficient with the hold-to-maturity asset share (hold-to-maturity assets
   as a share of total assets) measured in 2022q4. Column (3) reports unrealized hold-to-maturity losses scaled by tier 1 capital. Column (4) combines Columns (1)-(3). Column (5) adds the interaction of uninsured deposit share with HTM Asset Share and Unrealized HTM Losses. All variables (except for the cumulative returns) are mean zero and standarized to have standard deviation one, prior to interactions.}
   \bigskip
   \centering
   \begin{adjustbox}{width = 1.2\textwidth, center}
      \begin{tabular}{lccccc}
         \toprule
                                                                                  & (1)            & (2)            & (3)            & (4)            & (5)\\  
         \midrule 
         Constant                                                                 & -0.156$^{***}$ & -0.156$^{***}$ & -0.156$^{***}$ & -0.156$^{***}$ & -0.145$^{***}$\\   
                                                                                  & (0.008)        & (0.008)        & (0.008)        & (0.007)        & (0.007)\\   
         Uninsured Deposit Share                                                  & -0.048$^{***}$ &                &                & -0.035$^{***}$ & -0.030$^{***}$\\   
                                                                                  & (0.015)        &                &                & (0.011)        & (0.009)\\   
         HTM Asset Share                                                          &                & -0.046$^{**}$  &                & -0.020         & -0.008\\   
                                                                                  &                & (0.018)        &                & (0.015)        & (0.007)\\   
         Unrealized HTM Losses / Tier 1 Capital                                   &                &                & -0.062$^{***}$ & -0.048$^{***}$ & 0.007\\   
                                                                                  &                &                & (0.014)        & (0.009)        & (0.018)\\   
         Uninsured Deposit Share $\times$ HTM Asset Share                         &                &                &                &                & -0.008\\   
                                                                                  &                &                &                &                & (0.008)\\   
         Uninsured Deposit Share $\times$ Unrealized HTM Losses / Tier 1 Capital  &                &                &                &                & -0.040$^{***}$\\   
                                                                                  &                &                &                &                & (0.015)\\   
          \\
         Observations                                                             & 224            & 224            & 222            & 222            & 222\\  
         R$^2$                                                                    & 0.139          & 0.124          & 0.224          & 0.328          & 0.420\\  
         Adjusted R$^2$                                                           & 0.135          & 0.120          & 0.221          & 0.318          & 0.406\\  
         \bottomrule
      \end{tabular}
   \end{adjustbox}
\end{table}

\subsubsection{What was the same this time and what differed? }
\label{sec:whatwasdifferent}

The likelihood of depositors' panic decreases when banks possess more liquid assets, more capital, or better-performing assets. Liquid assets, including cash and securities, can function as liquidity buffers to meet withdrawals, which helps to alleviate depositors' concerns as well as losses from disorderly liquidations of illiquid assets (\textcite{BIS13}). Higher capital buffers also mitigate depositors' concerns by providing more loss absorption capacity. Furthermore, depositors have less reason to panic and withdraw when banks' assets are safe and expected to generate greater future returns. We next investigate whether these factors contributed to the spillover effects associated with the failure of \ac{SVB}.

\paragraph{Liquid assets}

We begin by examining the role of liquid assets, consisting of both cash and securities. We then analyze the effects of cash and securities separately to assess whether they had differential impacts.
 
 We report the  linear regression estimates of these relationships in Table \ref{tab:liquid_assets}, where the dependent variable is banks' early excess returns as in Table \ref{tab:dep_htm}.  In Column 1, the results indicate that overall holdings of liquid assets are not significantly associated with banks' performance in response to the shock from the bank run, which differs from the typical implications. We see the corresponding graphical relationship in Panel A of \Cref{fig:returns_v_liquid_assets} in the Appendix. In fact, despite \ac{SVB} holding more than 60\% of its assets in cash and securities, it failed to withstand deposit withdrawals. Unlike typical bank run scenarios, depositor withdrawals exacerbated financial distress not through fire-sale losses of illiquid assets, but rather by realizing accounting losses on highly liquid securities that were marked as held-to-maturity. This finding holds  even when controlling for banks' uninsured deposit reliance, as shown in column 3.  

However, not all liquid assets yielded similar results. In Column 2, we find that higher cash holdings were indeed associated with a significant mitigation of the negative shock, whereas this was not the case with securities. HTM securities, despite being liquid, proved problematic in this specific  episode as their liquidation would lead to the recognition of unrealized losses, thereby exacerbating the distress.  We see the corresponding graphical relationship in Panel B and C of \Cref{fig:returns_v_liquid_assets}. 

The benefit of holding cash becomes more pronounced when we control for funding outflow risks by incorporating uninsured deposit shares in column 4. This results in a significant increase in the explanatory power (both for the estimate's economic and statistically significance and $R^2$).  The estimate indicates that  a one-standard-deviation increase in cash to total assets would lead to a 3.2 percentage point increase in excess stock returns. In column 5, we additionally introduce an interaction term between cash (or securities) and uninsured deposit share. The estimate of the interaction term for cash is positive and significant, suggesting cash buffers effectively alleviate spillovers, particularly for banks facing higher funding outflow risks.

\begin{table}[htbp]
   \caption{\label{tab:liquid_assets} Cumulative returns correlated with liquid assets. This table reports estimated coefficients of regressions with the cumulative returns in excess of the S\&P 500 index for the banks in our sample from February 1, 2023 to March 17, 2023 as the outcome. In Column (1), we report the bivariate relationship with the liquid asset share (securities + cash scaled by total
   assets) measured in 2022q4. Column (2) adds uninsured deposit share to Column (1). Column (3) reports the coefficient with the cash scaled by total assets and securities scaled by total assets. Column (4) adds uninsured deposit share to Column (3). Column (5) interacts uninsured deposit share with cash and securities. All variables (except for the cumulative returns) are mean zero and standarized to have standard deviation one, prior to interactions.}
   \bigskip
   \centering
   \begin{adjustbox}{width = 1.2\textwidth, center}
      \begin{tabular}{lccccc}
         \toprule
                                                                     & (1)            & (2)            & (3)            & (4)            & (5)\\  
         \midrule 
         Constant                                                    & -0.156$^{***}$ & -0.156$^{***}$ & -0.156$^{***}$ & -0.156$^{***}$ & -0.157$^{***}$\\   
                                                                     & (0.009)        & (0.009)        & (0.008)        & (0.008)        & (0.008)\\   
         Liquid Assets / Total Assets                                & -0.009         &                & 0.007          &                &   \\   
                                                                     & (0.012)        &                & (0.011)        &                &   \\   
         Cash / Total Assets                                         &                & 0.015$^{**}$   &                & 0.032$^{***}$  & 0.024$^{***}$\\   
                                                                     &                & (0.007)        &                & (0.009)        & (0.008)\\   
         Securities / Total Assets                                   &                & -0.020         &                & -0.008         & -0.005\\   
                                                                     &                & (0.013)        &                & (0.011)        & (0.008)\\   
         Uninsured Deposit Share                                     &                &                & -0.051$^{***}$ & -0.056$^{***}$ & -0.054$^{***}$\\   
                                                                     &                &                & (0.015)        & (0.015)        & (0.014)\\   
         Cash / Total Assets $\times$ Uninsured Deposit Share        &                &                &                &                & 0.015$^{**}$\\   
                                                                     &                &                &                &                & (0.007)\\   
         Securities / Total Assets $\times$ Uninsured Deposit Share  &                &                &                &                & -0.015\\   
                                                                     &                &                &                &                & (0.015)\\   
          \\
         Observations                                                & 224            & 224            & 224            & 224            & 224\\  
         R$^2$                                                       & 0.005          & 0.040          & 0.142          & 0.202          & 0.240\\  
         Adjusted R$^2$                                              & 0.0004         & 0.031          & 0.134          & 0.191          & 0.223\\  
         \bottomrule
      \end{tabular}
   \end{adjustbox}
\end{table}

\begin{figure}[thbp]
    \centering
    \includegraphics[width=\linewidth]{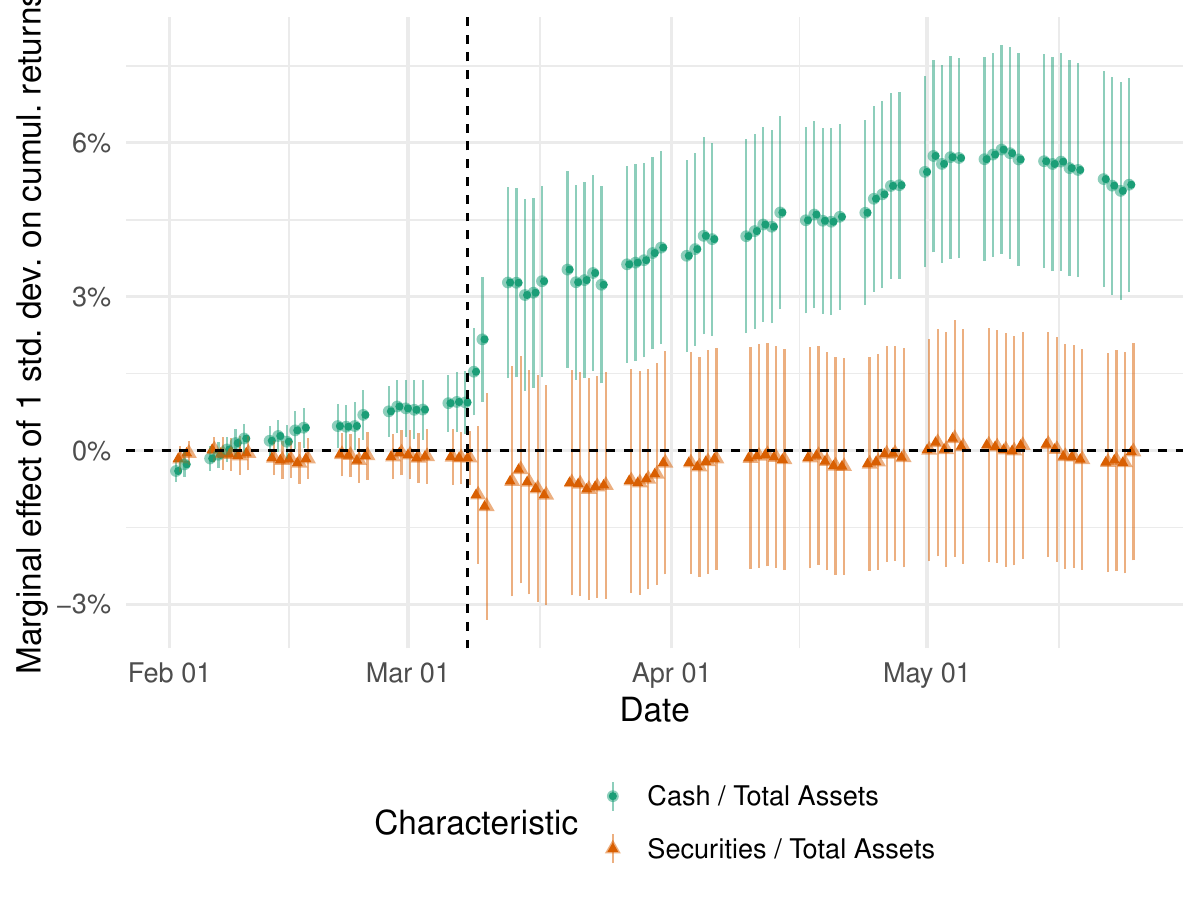}
    \caption{Cumulative returns correlated with liquid assets, over time. This graph plots the coefficients for cash and securities (Column 4 of \Cref{tab:liquid_assets}) for every trading day, holding fixed uninsured deposit share. We omit day fixed effects, which correspond to the constant for each cross-sectional regression. }
    \label{fig:returns_coef_over_time_cash}
\end{figure}

Figure \ref{fig:returns_coef_over_time_cash} illustrates the dynamic association between banks' stock returns and different sources of liquid assets (cash or securities), as we did in 
Figure \ref{fig:returns_coef_over_time}. Prior to the run, banks' cash holdings slowly began to predict higher cumulative returns across banks, but securities did not. However, following the run, the impact of cash holdings jumped immediately and more after the initial weekend. This effect becoming more pronounced over time. In contrast, banks' securities holdings did not significantly affect their stock returns throughout the sample period.

\paragraph{Capitalization}
Next, we examine whether banks' capitalization, as measured by the ratio of Tier 1 capital to risk-weighted assets (RWA), played a role in mitigating the spillover effects.  As shown in \Cref{fig:returns_v_tier1capitalratio}, we observe that, consistent with typical banking crises, higher capital levels were associated with relatively higher returns, especially among undercapitalized banks. It is worth noting that \ac{SVB} does not appear to be specifically undercapitalized in this figure; but one should note that this capitalization measure does not account for the unrealized HTM losses.

We can confirm this relationship from the regression result reported in column 1 of Table \ref{tab:npl_capital}, where a one-standard-deviation increase in the Tier 1 capital ratio is associated with a 3.2 percentage point increase in excess early returns. The significance of capitalization in mitigating spillovers persists even when controlling for funding outflow risks, as shown in column 4, although the effect is less pronounced both statistically and economically. In column 5, we include the interaction terms to examine whether the stability effect of capitalization is more pronounced for banks facing higher funding outflow risks.
While the coefficient for the interaction of capitalization and uninsured deposit shares is positive and sizable, its statistical significance is somewhat marginal, with a $t$-statistic of 1.3.

\begin{table}[htbp]
   \caption{\label{tab:npl_capital} Cumulative returns correlated with NPL and Tier 1 capital. This table reports estimated coefficients of regressions with the cumulative returns in excess of the S\&P 500 index for the banks in our sample from February 1, 2023 to March 17, 2023 as the outcome. %
      In Column (1), we report the bivariate relationship with 
      the tier 1 capital ratio measured in 2022q4. %
      Column (2) reports the coefficient with the non-performing loan ratio (non-performing loans scaled by total loans) measured in 2022q4.
      Column (3) combines Column (1) and (2). %
       Column (4) adds uninsured deposit share to Column (3). %
       Column (5) interacts uninsured deposit share with non-performing loans and tier 1 capital. %
       All variables (except for the cumulative returns) are mean zero and standarized to have standard deviation one, prior to interactions.}
   \bigskip
   \centering
   \begin{adjustbox}{width = 1.2\textwidth, center}
      \begin{tabular}{lccccc}
         \toprule
                                                                              & (1)            & (2)            & (3)            & (4)            & (5)\\  
         \midrule 
         Constant                                                             & -0.159$^{***}$ & -0.156$^{***}$ & -0.159$^{***}$ & -0.158$^{***}$ & -0.151$^{***}$\\   
                                                                              & (0.009)        & (0.009)        & (0.009)        & (0.008)        & (0.009)\\   
         Tier 1 Capital Ratio                                                 & 0.032$^{***}$  &                & 0.032$^{***}$  & 0.022$^{**}$   & 0.036$^{*}$\\   
                                                                              & (0.011)        &                & (0.011)        & (0.011)        & (0.019)\\   
         Non-Performing Loans / Total Loans                                   &                & 0.003          & 0.0008         & -0.0003        & -0.004\\   
                                                                              &                & (0.009)        & (0.009)        & (0.008)        & (0.008)\\   
         Uninsured Deposit Share                                              &                &                &                & -0.047$^{***}$ & -0.049$^{***}$\\   
                                                                              &                &                &                & (0.017)        & (0.016)\\   
         Tier 1 Capital Ratio $\times$ Uninsured Deposit Share                &                &                &                &                & 0.032\\   
                                                                              &                &                &                &                & (0.024)\\   
         Non-Performing Loans / Total Loans $\times$ Uninsured Deposit Share  &                &                &                &                & 0.002\\   
                                                                              &                &                &                &                & (0.012)\\   
          \\
         Observations                                                         & 216            & 224            & 216            & 216            & 216\\  
         R$^2$                                                                & 0.060          & 0.0004         & 0.060          & 0.173          & 0.221\\  
         Adjusted R$^2$                                                       & 0.056          & -0.004         & 0.051          & 0.161          & 0.202\\  
         \bottomrule
      \end{tabular}
   \end{adjustbox}
\end{table}

\begin{figure}[thbp]
    \centering
    \includegraphics[width=\linewidth]{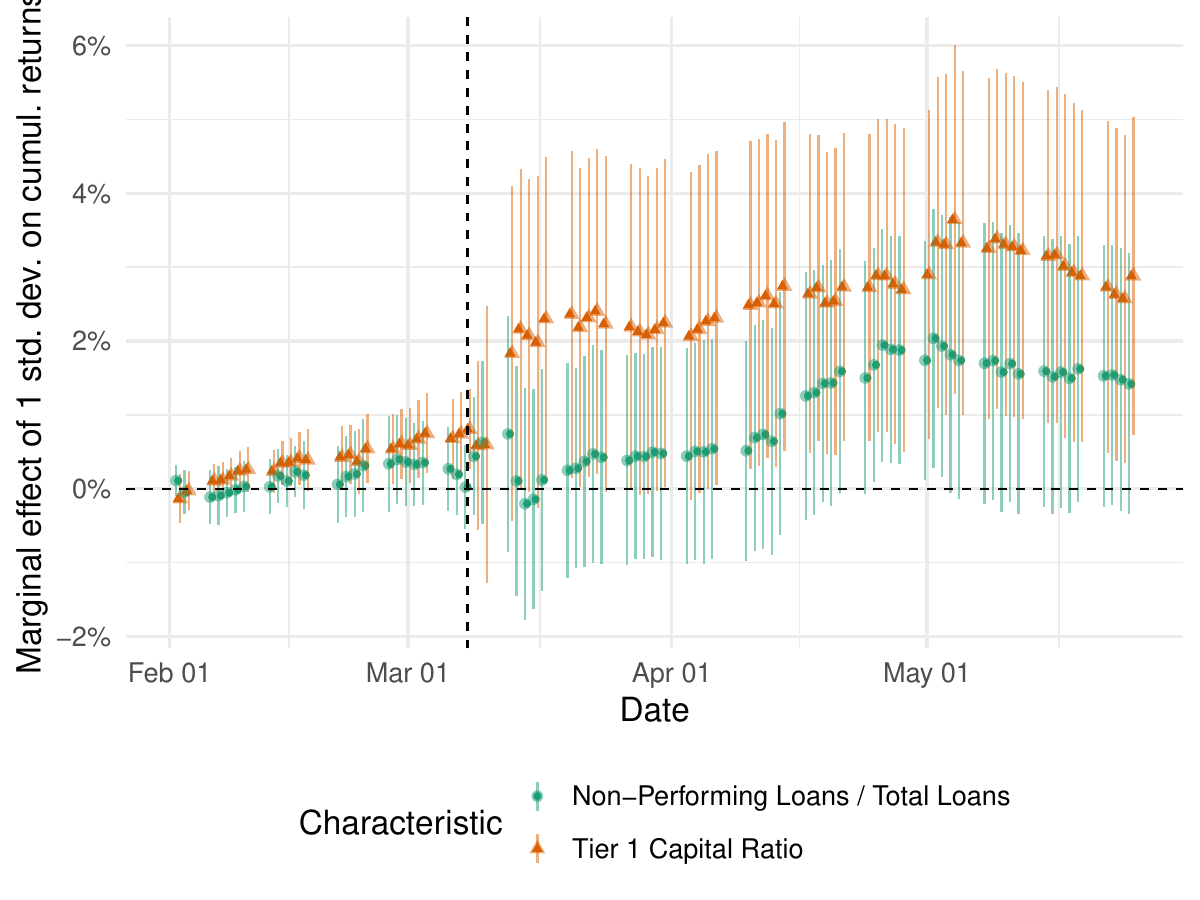}
    \caption{Cumulative returns correlated with NPL and Tier 1 capital, over time. This graph plots the coefficients for NPL and Tier 1 Capital ratios (Column 4 of \Cref{tab:npl_capital}) for every trading day, holding fixed uninsured deposit share. We omit day fixed effects, which correspond to the constant for each cross-sectional regression. }
    \label{fig:returns_coef_over_time_npl}
\end{figure}

In Figure \ref{fig:returns_coef_over_time_npl}, we plot the dynamic correlations of the factors over time. Through February, bank capitalization has a growing positive correlation with cumulative returns, and this effect jumps significantly following the weekend after the SVB run.

\paragraph{Asset quality}

Finally, we examine whether asset quality, as measured by the non-performing loan (NPL) ratio, contributed the contagion effect. In typical banking crises, concerns about bank asset quality, or ``fundamentals,'' can exacerbate depositor panic and bank runs (\cite{rochet2004coordination,goldstein2005demand}). However, in the case of the 2023 crisis, depositor panic stemmed from implied losses due to duration mismatch rather than credit risk exposures. As illustrated in \Cref{fig:returns_vs_NPL}, there was no discernible effect of better-performing assets in mitigating the negative spillovers. Moreover, the regression estimates in Table \ref{tab:npl_capital} indicate that the NPL ratio was not a statistically significant predictor in all of the specifications. The null effect is also confirmed in Figure   \ref{fig:returns_coef_over_time_npl} displaying the dynamic impact.

\bigskip

In sum, asset liquidity in general or better-performing assets did not help mitigate the spillovers during the bank runs of 2023. Securities like Treasury or government-sponsored enterprise (GSE) bonds, while liquid, did not reduce bank stress because their liquidations would result in realizations of implied losses. Cash and capital, on the other hand, were still effective in reducing the stress. It was uninsured deposits that became problematic, in combination with the implied losses in the HTM securities portfolio.  

\subsubsection{Did size matter? Immediate vs. medium-term effects}
\label{subsec:size}

\begin{figure}[thbp]
    \centering
    \begin{subfigure}[b]{0.75\linewidth}
    \includegraphics[width=\linewidth]{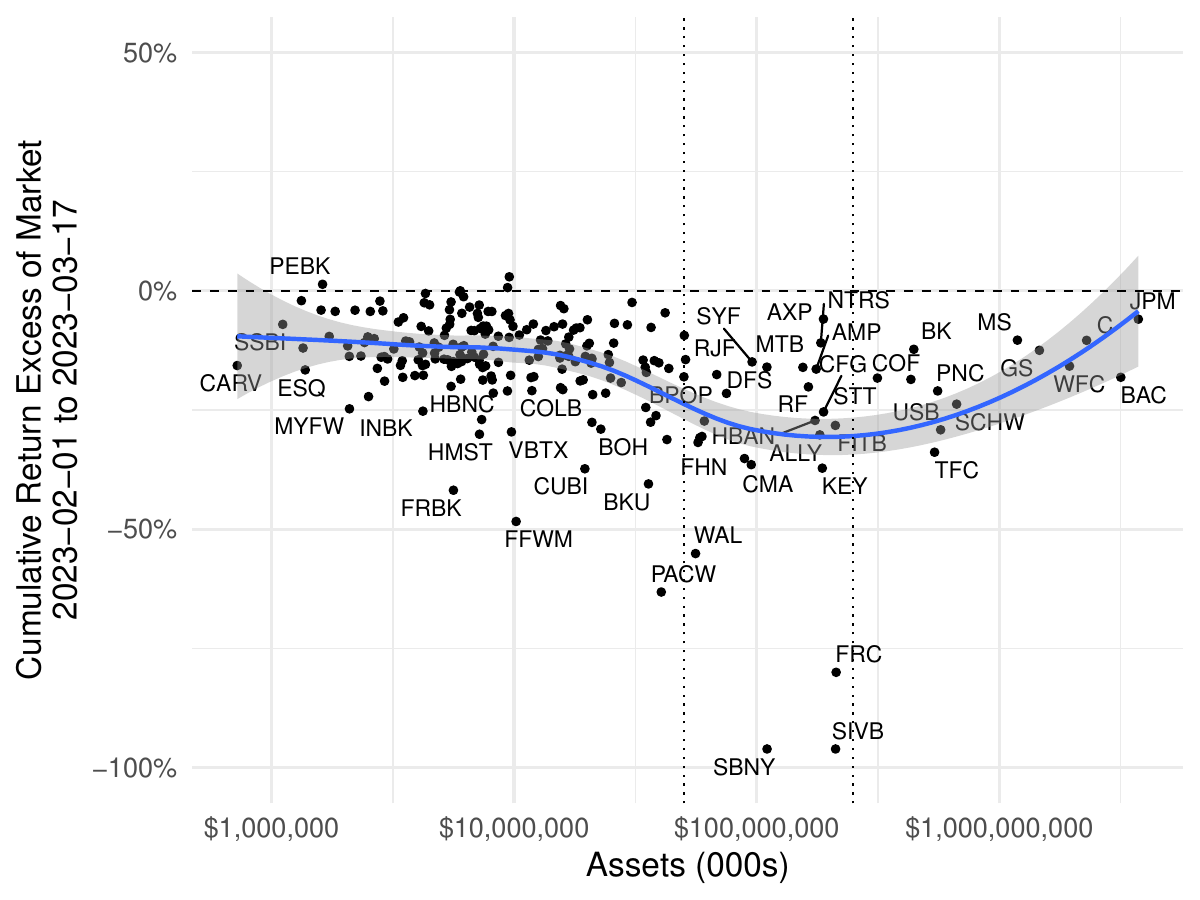}
        \caption{Early cumulative returns}
    \label{fig:returns_v_assets_early}
    \end{subfigure}
    \begin{subfigure}[b]{0.75\linewidth}
    \includegraphics[width=\linewidth]{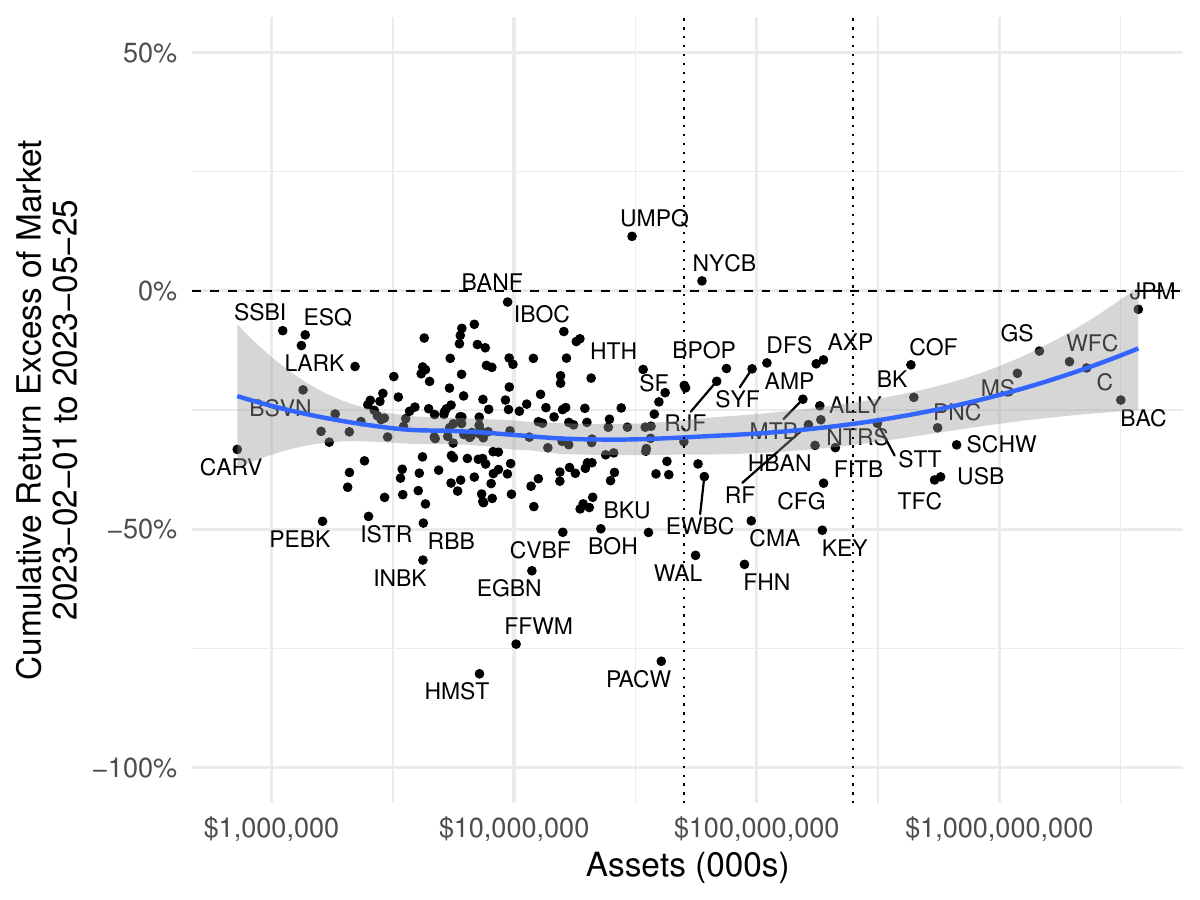}
        \caption{Late cumulative returns}
        \label{fig:returns_v_assets_late}
    \end{subfigure}
    \caption{Cumulative returns explained by total assets. Panel (a) of this figure plots the cumulative returns in excess of the S\&P 500 index for the banks in our sample from February 1, 2023 to March 17, 2023 against the assets measured in 2022q4. Panel (b) of this figure plots the cumulative returns in excess of the S\&P 500 index for the banks in our sample from February 1, 2023 to May 25, 2023 against the assets measured in 2022q4. See Section \ref{sec:data} for details on the data construction.  The fitted lines are local linear regression (loess) curves. Stocks are denoted by their ticker symbol; Silicon Valley Bank's ticker is SIVB. }
    \label{fig:returns_v_assets}
\end{figure}

Next, we investigate whether bank size was  associated with negative returns. Figure \ref{fig:returns_v_assets_early} displays the ``early'' cumulative returns plotted against bank assets. The figure suggests the presence of differential spillover effects across different asset size categories. A notable observation is the heightened stress experienced by mid-sized banks, particularly super-regional banks falling between assets of \$50 billion and \$250 billion, denoted by the two vertical lines. As discussed in Section 2, these banks, including \ac{SVB}, were initially designated as systemically important following the regulatory reforms implemented after the GFC, but  faced relaxed regulations with the regulatory rollback in 2018. One possible interpretation is that investors exhibited greater concerns with banks of a similar size to \ac{SVB}, particularly in the aftermath of its failure.\footnote{This is consistent with the findings in \textcite{LuckPlosserYounger2023}, indicating the deposit outflows were most acute among the super-regional banks, with some funds flowing into the larger banks.}

\begin{table}[htbp]
   \caption{\label{tab:dep_assets} Cumulative returns correlated with asset size. This table reports estimated coefficients of regressions with the cumulative returns in excess of the S\&P 500 index for the banks in our sample. 
      In Column (1), we report the relationship with binned indicator variables of total assets measured in 2022q4. The bins exhaustively bin out all observations, and do not include a constant, so each coefficient is the average for each bin. 
      Column (2) includes the binned controls for assets (excluding the bin for banks with total assets less than 5 billion dollars) and a constant, as well as the controls for uninsured deposit share, tier 1 capital ratio, cash/ total assets and unrealized hold-to-maturity losses. We also include a control for the individual banks' estimated factor loading (beta) on the bank index in excess of the S\&P500 in 2022. This captures any systematic loading on overall bank movements. Column 3 adds the cumulative abnormal return from February 1, 2022 to January 31, 2023 to Column (2). Columns (1)-(3) use cumluative returns from February 1, 2023 to March 17, 2023 as the outcome. Columns (4)-(6) repeat Columns (1)-(3) and use cumluative returns from February 1, 2023 to May 25, 2023 as the outcome.
       All variables (except for the cumulative returns) are mean zero and standarized to have standard deviation one, prior to interactions.}
   \bigskip
   \centering
   \begin{adjustbox}{width = 1.2\textwidth, center}
      \begin{tabular}{lcccccc}
         \toprule
                                                 & (1)            & (2)            & (3)            & (4)            & (5)            & (6)\\  
         \midrule 
         Assets [0-5b]                           & -0.113$^{***}$ &                &                & -0.291$^{***}$ &                &   \\   
                                                 & (0.009)        &                &                & (0.015)        &                &   \\   
         Assets (5b-10b]                         & -0.115$^{***}$ & 0.032$^{**}$   & 0.027$^{*}$    & -0.287$^{***}$ & 0.021          & 0.016\\   
                                                 & (0.010)        & (0.016)        & (0.016)        & (0.015)        & (0.029)        & (0.027)\\   
         Assets (10b-50b]                        & -0.160$^{***}$ & 0.005          & -0.013         & -0.318$^{***}$ & 0.003          & -0.016\\   
                                                 & (0.013)        & (0.020)        & (0.020)        & (0.017)        & (0.036)        & (0.033)\\   
         Assets (50b-250b]                       & -0.314$^{***}$ & -0.110$^{***}$ & -0.119$^{***}$ & -0.335$^{***}$ & 0.010          & 0.0009\\   
                                                 & (0.046)        & (0.039)        & (0.042)        & (0.063)        & (0.076)        & (0.079)\\   
         Assets (250b-1tr]                       & -0.224$^{***}$ & -0.045         & -0.048         & -0.293$^{***}$ & 0.040          & 0.037\\   
                                                 & (0.026)        & (0.037)        & (0.038)        & (0.031)        & (0.058)        & (0.059)\\   
         Assets (1tr-10tr]                       & -0.122$^{***}$ & 0.089$^{**}$   & 0.069          & -0.146$^{***}$ & 0.205$^{***}$  & 0.183$^{***}$\\   
                                                 & (0.016)        & (0.043)        & (0.043)        & (0.024)        & (0.065)        & (0.063)\\   
         Constant                                &                & -0.082$^{***}$ & -0.114$^{***}$ &                & -0.301$^{***}$ & -0.334$^{***}$\\   
                                                 &                & (0.020)        & (0.019)        &                & (0.033)        & (0.033)\\   
         Uninsured Deposit Share                 &                & -0.034$^{***}$ & -0.025$^{**}$  &                & -0.064$^{***}$ & -0.054$^{***}$\\   
                                                 &                & (0.013)        & (0.011)        &                & (0.014)        & (0.013)\\   
         Tier 1 Capital Ratio                    &                & 0.005          & 0.008          &                & 0.006          & 0.009\\   
                                                 &                & (0.009)        & (0.008)        &                & (0.011)        & (0.011)\\   
         Cash / Total Assets                     &                & 0.036$^{***}$  & 0.023$^{***}$  &                & 0.048$^{***}$  & 0.034$^{***}$\\   
                                                 &                & (0.009)        & (0.009)        &                & (0.011)        & (0.010)\\   
         Unrealized HTM Losses / Tier 1 Capital  &                & -0.046$^{***}$ & -0.042$^{***}$ &                & -0.040$^{***}$ & -0.036$^{***}$\\   
                                                 &                & (0.009)        & (0.007)        &                & (0.009)        & (0.009)\\   
         Beta on Bank Index (Excess of S\&P500)  &                & -0.124$^{***}$ & -0.070         &                & -0.027         & 0.031\\   
                                                 &                & (0.047)        & (0.047)        &                & (0.082)        & (0.081)\\   
         Cumulative Abnormal Returns (2022)      &                &                & 0.254$^{***}$  &                &                & 0.269$^{***}$\\   
                                                 &                &                & (0.060)        &                &                & (0.094)\\   
          \\
         Observations                            & 224            & 216            & 216            & 224            & 216            & 216\\  
         R$^2$                                   & 0.242          & 0.545          & 0.613          & 0.032          & 0.315          & 0.363\\  
         Adjusted R$^2$                          & 0.225          & 0.523          & 0.592          & 0.010          & 0.282          & 0.329\\  
         \bottomrule
      \end{tabular}
   \end{adjustbox}
\end{table}

The regression estimates presented in Table \ref{tab:dep_assets} provide further confirmation of the particular stress experienced by these regional banks with similar size immediately after the \ac{SVB} failure. In Column 1, dummy variables are included for each of the six size buckets, where each estimate indicates the average ``early'' excess returns for that specific size group relative to the market returns. For only this column, we exclude the constant, so each coefficient should represent the average return in the size bucket. Column 2 incorporates the bank characteristics (i.e., vulnerability factors) that we identified as being associated with the spillovers, including uninsured deposit share, capitalization, cash holdings, and unrealized HTM security losses. Here, a constant term is included and we exclude the size indicator variable for the smallest asset size group, such that the estimates for the asset groups reflect excess returns relative to banks with less than 5 billion dollars in assets. Finally, we include ``bank-beta,'' which is defined as the individual banks' factor loading on the Dow Jones U.S. Banks Index net of the S\&P 500 index, estimated for stock returns in 2022. This variable accounts for any systemic loading on the overall banking sector performance, assessed prior to the SVB failure.        

The estimates in columns 1 and 2 highlight the significant ``early'' stress experienced by the super-regional banks. When compared to banks with assets below \$5 billion, banks within the size range of \$50 billion and \$250 billion experienced significantly lower returns, with returns that were 11 percentage points lower conditional on observables, and 20 percentage points lower unconditionally. Interestingly, conditional on observables, the largest banks with assets greater than \$1 trillion showed relative outperformance compared to smaller banks.

The estimates in column 2 indicate that unrealized HTM losses were the most economically significant predictors of the scale of spillover effects immediately after the \ac{SVB} failure, followed by uninsured deposits and cash. These correlations held even after controlling for each bank's bank-beta, which itself had a negative effect. This suggests that banks that performed worse likely did load more on an overall banking sector factor, but the other observables explained additional variation. 

\textbf{Differential medium-term effect} The size effect, however, presents a different pattern over time. Figure \ref{fig:returns_v_assets_late} displays banks' ``late'' returns two months after the \ac{SVB} failure, from February 1, 2023 to May 25, 2023. Unlike the immediate effect shown in Figure \ref{fig:returns_v_assets_early}, the substantial negative impact had spread beyond mid-sized banks in the \$50 billion and \$250 billion range to include smaller banks as well. 

The regression estimates reported in columns 4 and 5 of Table 5  repeats the exercise from column 1 and 2 for the late returns. The banking sector, as a whole, experienced a further decline in stock prices over time (column 4) compared to immediately after the \ac{SVB} failure (column 1). By late May, all banks had experienced negative returns of nearly 30\%, except for the largest banks.  While policy measures, such as the FDIC's full protection of failed bank deposits and the Federal Reserve's new lending facility, helped prevent additional bank runs, concerns about the soundness of the entire banking system spread. This is potentially because the fundamental driver of banking stress was the unrealized losses in banks' long-term assets, which included not only HTM securities but also fixed-rate mortgage loans, which affected a broad range of banks.\footnote{Assessing the scale of banks' mark-to-market losses encompassing both securities and mortgage loans, \textcite{jiang2023} argue that the loss for the U.S. banking system amounts to \$2 trillion and is widespread. Also see \textcite{FlannerySorescu2023} providing similar assessments.} The absence of any indication of a rate cut by the central bank further heightened these concerns, resulting in stress for the entire banking sector. 

However, amidst broader stress, we can observe even stronger outperformance of the largest banks compared to the rest of the system. In column 5, banks with assets greater than \$1 trillion exhibited returns 20 percentage points higher than the smallest banks, more than twice the difference observed in column 2. Conversely, none of the other size groups performed differently. This trend aligns with the earlier observations of equal-weighted and value-weighted returns.

\begin{figure}[thbp]
    \centering
    \includegraphics[width=\linewidth]{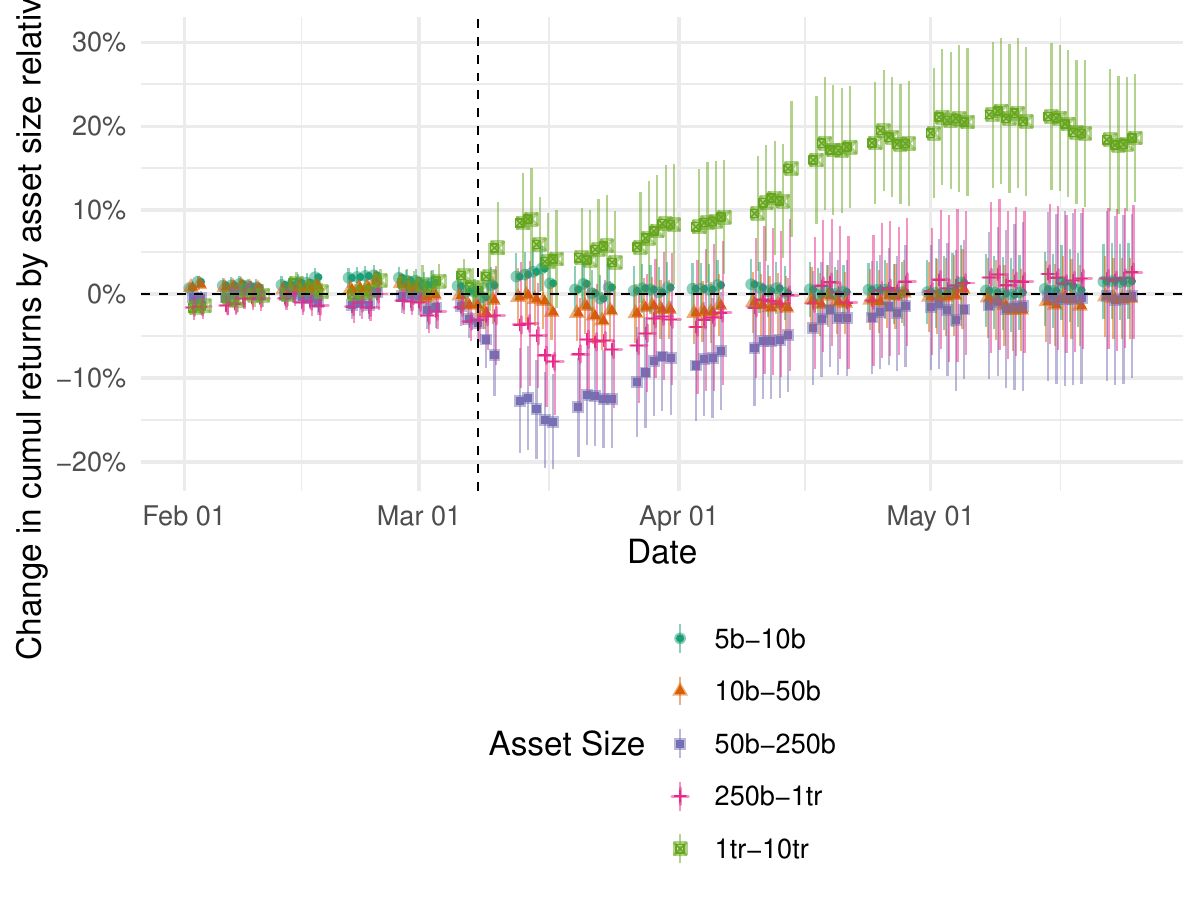}
    \caption{Cumulative returns correlated with asset size, over time. This graph plots the cumulative return by asset size bin (Column 2 of \Cref{tab:dep_assets}) for every trading day, relative to the [0-5b] size category, and controlling for the other observables in Column 2. }
    \label{fig:returns_coef_over_time_size}
\end{figure}

Figure \ref{fig:returns_coef_over_time_size} clearly illustrates this dynamic trend. Following the SVB failure, we begin to observe specific spillovers affecting banks with assets between \$50 billion and \$250 billion, relative to the baseline banks with assets smaller than \$5 billion. However, this differential impact dissipates over time. On the other hand, the largest banks, with assets greater than \$1 trillion, did not significantly outperform others at the onset of the crisis, but gradually began to do so over time.

Why did investors treat these largest banks differently during a period of systemic stress? Evidently, these banks were subject to more robust regulation and supervision than small banks, including \ac{SVB}, and thus were likely in better financial health. However, such regulatory difference alone does not fully explain their outperformance. Note that relatively smaller large banks, with assets greater than \$250 billion but smaller than \$1 trillion, were also designated as ``systemically important,'' to face similarly strict regulations.  However, these banks did not outperform the rest -- only those larger than \$1 trillion did so.
 
One possible explanation for this pattern is the implicit guarantees associated with too-big-to-fail (TBTF) banks (\textcite{OharaShaw1990}, \textcite{PenasUnal2004}, \textcite{BaronSchularickZimmermann2023}). Even among the similarly regulated ``systemically important financial institutions'' larger than \$250 billion in assets, investors seemed to have priced in greater TBTF benefits for the largest banks. In fact, being perceived as safe, these banks also experienced deposit inflows amid a flight to quality by depositors, which further helped them  mitigate the system-wide stress (\textcite{caglio2023}).\footnote{Examining historical data since 1870 across 17 advanced economies, \textcite{BaronSchularickZimmermann2023} found that during banking crises, regulators are significantly more inclined to rescue their largest (i.e., top-5) banks, and these banks' deposits exhibit greater stability.} Those with deposit outflows, on the other hand, were forced to resort to more expensive funding sources, such as time deposits and borrowing from the Federal Reserve or Federal Home Loan Banks (\textcite{KangLuckPlosser2023}).

\subsection{What was expected by the market and what was a surprise?}

Having established the factors associated with the spillovers, our next focus is to examine whether market participants had anticipated these vulnerabilities in advance or if they were caught by surprise following the \ac{SVB} bank run. As discussed in Section 4.1, there was a strong association between banks' stock returns in 2022 and those immediately after the \ac{SVB} bank run, indicating investors had expectations of vulnerabilities before the run occurred.

To assess this issue, we include each bank's 2022 return as an additional control for the regression specification of Table \ref{tab:dep_assets} and present the estimates in columns 3 and 6. Focusing on the early returns analyzed in column 3, note that bank-beta, which accounts for market perceptions of systematic exposures to the banking industry performance, no longer predicts the spillovers; only 2022 returns do. This suggests that investors priced in certain vulnerabilities that are not particularly associated with typical industry factors.  

To explore the specific factors investors priced in, we compare estimates in columns 2 and 3, with the latter controlling for the 2022 return. We observe that vulnerabilities associated with uninsured deposit reliance or cash holdings were partly anticipated by investors, as indicated by their diminished significance when controlling for the 2022 returns. However, the inclusion of the lagged returns has little effect on the explanatory power of HTM losses or the size dummy for the super-regional banks (50b-250), whose magnitude, in fact becomes larger. This indicates that investors were potentially surprised by these factors, leading to additional negative returns in response to the \ac{SVB} bank run. Comparison of columns 5 and 6 for the late returns indicates that investors did not anticipate the ``TBTF'' benefit for the largest banks (greater than \$1 trillion), either. 

To further explore this, we evaluate the correlation between each bank's stock return prior to the 2023 crisis (i.e., ``2022 returns'' from 2022-02-01 to 2023-01-31) and their vulnerability characteristics. A stronger correlation would indicate that investors had more pronounced concerns regarding specific vulnerabilities. Note that we hold these vulnerability factors fixed to the observed measures from the 2022q4 filings.

In Table \ref{tab:dep_2022}, we regress banks' 2022 returns on the respective vulnerability factors we have previously assessed. Focusing on the factors we identified to be relevant in Table \ref{tab:dep_assets}, our results indicate that market investors to some degree accounted for possible losses associated with heavy reliance on uninsured deposits and limited cash holdings, as evidenced by their significant correlation with the 2022 returns. These losses could result from increases in funding costs or deposit outflows in the face of central bank tightening.\footnote{See, e.g., \textcite{DrechslerSavovSchnablWang2023} or \textcite{KangLuckPlosser2023}.} Furthermore, the inclusion of these factors as regression controls leads to a distinct increase in the $R^2$ value. Notably, the vulnerabilities associated with asset size (both the specific negative impact on super-regional banks and the positive impact on the largest banks) and HTM unrealized losses did not appear to be reflected in the 2022 returns. This indicates that investors considered the recognition of implied securities loss, which would materialize only if banks were forced to sell these securities, to be unlikely.

\begin{table}[htbp]
   \caption{\label{tab:dep_2022} Cumulative returns in 2022 correlations. This table reports estimated coefficients of regressions with the cumulative returns in excess of the S\&P 500 index for the banks in our sample from February 1, 2022 to January 31, 2023. 
      In Column (1), we report the relationship with binned indicator variables of total assets measured in 2022q4. The bins exhaustively bin out all observations, and do not include a constant, so each coefficient is the average for each bin. 
      Column (2) includes the binned controls for assets (excluding the bin for banks with total assets less than 5 billion dollars) and a constant, as well as the controls for uninsured deposit share, hold-to-maturity asset share, and unrealized hold-to-maturity losses.
      Column (3) includes  cash/ total assets and securities / total assets.
      Column (4) includes  tier 1 capital ratio and non-performing loan ratio.
       All variables are mean zero and standarized to have standard deviation one, prior to interactions.}
   \bigskip
   \centering
   \begin{adjustbox}{width = 1\textwidth, center}
      \begin{tabular}{lcccc}
         \toprule
                                                & (1)           & (2)          & (3)            & (4)\\  
         \midrule 
         Assets [0-5b]                          & 0.050$^{**}$  &              &                &   \\   
                                                & (0.024)       &              &                &   \\   
         Assets (5b-10b]                        & 0.023         & -0.025       & -0.017         & -0.025\\   
                                                & (0.019)       & (0.031)      & (0.030)        & (0.029)\\   
         Assets (10b-50b]                       & 0.043$^{***}$ & 0.005        & 0.010          & 0.005\\   
                                                & (0.015)       & (0.029)      & (0.029)        & (0.029)\\   
         Assets (50b-250b]                      & -0.006        & -0.039       & -0.060         & -0.060\\   
                                                & (0.037)       & (0.042)      & (0.041)        & (0.041)\\   
         Assets (250b-1tr]                      & -0.038$^{*}$  & -0.061       & -0.085$^{**}$  & -0.084$^{**}$\\   
                                                & (0.022)       & (0.043)      & (0.038)        & (0.038)\\   
         Assets (1tr-10tr]                      & 0.005         & -0.008       & -0.033         & -0.018\\   
                                                & (0.038)       & (0.045)      & (0.041)        & (0.046)\\   
         Constant                               &               & 0.041        & 0.040          & 0.046$^{*}$\\   
                                                &               & (0.025)      & (0.025)        & (0.024)\\   
         Uninsured Deposit Share                &               & -0.021$^{*}$ & -0.034$^{***}$ & -0.042$^{***}$\\   
                                                &               & (0.012)      & (0.012)        & (0.013)\\   
         HTM Asset Share                        &               & -0.007       & -0.005         & -0.006\\   
                                                &               & (0.015)      & (0.013)        & (0.013)\\   
         Unrealized HTM Losses / Tier 1 Capital &               & -0.012       & -0.007         & -0.007\\   
                                                &               & (0.011)      & (0.012)        & (0.011)\\   
         Cash / Total Assets                    &               &              & 0.048$^{***}$  & 0.052$^{***}$\\   
                                                &               &              & (0.010)        & (0.011)\\   
         Securities / Total Assets              &               &              & -0.0005        & -0.0002\\   
                                                &               &              & (0.010)        & (0.011)\\   
         Tier 1 Capital Ratio                   &               &              &                & -0.003\\   
                                                &               &              &                & (0.012)\\   
         Non-Performing Loans / Total Loans     &               &              &                & -0.014\\   
                                                &               &              &                & (0.010)\\   
          \\
         Observations                           & 224           & 222          & 222            & 216\\  
         R$^2$                                  & 0.016         & 0.058        & 0.145          & 0.175\\  
         Adjusted R$^2$                         & -0.007        & 0.023        & 0.104          & 0.127\\  
         \bottomrule
      \end{tabular}
   \end{adjustbox}
\end{table}

How can we interpret these null results when all relevant information was publicly available?  The information view of banking panics (see, e.g., \textcite{Gorton1988},  \textcite{DangGortonHolmstrom2020}) can provide a plausible framework. The view argues
that abrupt shifts in depositors' risk perception of bank assets, triggered by the arrival of negative news such as a failure of a large institution or adverse macroeconomic perspectives, can cause widespread bank runs.

Even though the presence of unrealized losses in banks' securities portfolios affects their fundamentals, this vulnerability would only materialize when banks were forced to liquidate these assets to generate cash in the face of severe deposit outflows. 
Hence, unless depositors are concerned about these implied losses, stock investors have little reason to pay close attention to them, either. While investors did take into account the possible negative impacts associated with uninsured deposits, whose supply can be more sensitive to rate hikes, they at least did not anticipate the implied losses to be realized with substantial withdrawals. The SVB episode, however, potentially caused a significant shift in this dynamic through depositors' risk perceptions. 

A similar argument can be made for bank size. While the size of a bank can influence its likelihood of survival, this factor becomes critical  particularly when experiencing systemic instability. Following the \ac{SVB} failure, investors became aware of the vulnerability faced by mid-sized banks and the relaxation of regulations that they experienced. As concerns about the entire banking system grew over time, flying-to-quality investors sought safety in the largest ``systemic'' banks. However, these effects are significant only when conditioned on the occurrence of a systemic disruption, which may have a small unconditional probability ex ante. 

It is important to note that lack of information was not a constraint. The bank balance sheet information (including unrealized losses) had been announced for 2022q4 by the end of January 2023, and the accumulated HTM losses had been apparent prior to 2022q4 as well.\footnote{Assessing systemic crisis episodes between 1972 and 2022, \textcite{DuKarolyiLi2023} find that narrative-based accounting measures associated with specific crises, despite being used for post-crisis regulatory reforms, are less informative in predicting subsequent crises compared to   market-based measures.} Hence, there was no sense of hidden information on implied losses. Rather, the SVB bank run seems to have served as a device to change investors' risk perceptions to cause widespread spillovers.

\section{Conclusion}

This paper analyzes the \ac{SVB} bank failure, focusing on the bank balance sheet factors associated with contagion on the rest of the banking system.

Lessons from the GFC suggested that liquidity risks could be managed by holding liquid assets, including cash and ``high-quality’’ securities, or funding with deposits rather than wholesale funding. However, the \ac{SVB} episode showed that these measures may not always help prevent bank runs. The panic was triggered by concerns about unrealized losses in their HTM securities portfolio, albeit of high-quality, which led uninsured depositors to panic. This forced the banks to sell their HTM securities, resulting in the realization of implied losses, exacerbating the run. 

Furthermore, it is worth noting that, in this particular case, cash and liquid securities were not interchangeable in reducing damages from deposit outflows. While some banks had adequate HQLAs from the standpoint of the regulatory requirements such as \ac{LCR}, interest rate risks caused significant losses even on sufficiently liquid securities. These securities did not effectively serve as liquidity buffers against funding outflows; only cash and capital buffers were successful in absorbing losses and containing the spillovers.   

We also find that banks whose stocks experienced more negative returns following the failure of SVB had already underperformed in the previous year. Our finding suggests that investors, to some extent, anticipated the downsides associated with uninsured deposit reliance during the rate hikes.
 However, they did not foresee the damages associated with unrealized securities losses until the SVB run, even though the data was publicly available. This highlights the challenges for regulators in devising comprehensive and robust stress scenarios. 

In many respects, the SVB episode highlights new channels for bank failures and contagion, providing us with novel lessons. Some of the elegantly designed measures based on the experiences of the GFC did not effectively prevent the systemic instability this time. 
A major driver for the spillovers was implied losses in the banks’ securities portfolio, but this was not adequately considered in either the regulatory stress test scenarios or the investors’ risk pricing ex ante. 
Moreover, the treatment and incorporation of these implied losses into regulatory measures would pose important challenges as such losses crystallize only when banks are forced to sell held-to-maturity assets at market prices.  These are important topics for further research and, going forward, policymakers should incorporate these lessons into their policies to enhance the resilience of the financial system.

\pagebreak

\printbibliography%

@article{ivashina2010bank,
  title={Bank lending during the financial crisis of 2008},
  author={Ivashina, Victoria and Scharfstein, David},
  journal={Journal of Financial Economics},
  volume={97},
  number={3},
  pages={319--338},
  year={2010},
  publisher={Elsevier}
}

@article{hanson2015banks,
  title={Banks as patient fixed-income investors},
  author={Hanson, Samuel G and Shleifer, Andrei and Stein, Jeremy C and Vishny, Robert W},
  journal={Journal of Financial Economics},
  volume={117},
  number={3},
  pages={449--469},
  year={2015},
  publisher={Elsevier}
}

@article{bai2018measuring,
  title={Measuring Liquidity Mismatch in the Banking Sector},
  author={Bai, Jennie and Krishnamurthy, Arvind and Weymuller, Charles-Henri},
  journal={Journal of Finance},
  volume={73},
  number={1},
  pages={51--93},
  year={2018},
  publisher={Wiley Online Library}
}

@article{gatev2006banks,
  title={Banks' advantage in hedging liquidity risk: Theory and evidence from the commercial paper market},
  author={Gatev, Evan and Strahan, Philip E},
  journal={Journal of Finance},
  volume={61},
  number={2},
  pages={867--892},
  year={2006},
  publisher={Wiley Online Library}
}

@unpublished{BaronSchularickZimmermann2023,
  title={Survival of the biggest: Large banks and financial crises},
  author={Baron, Matthew and Schularick, Moritz and Zimmermann, Kaspar},
  	Note = {Working Paper},
  year={2023}
}

@unpublished{DuKarolyiLi2023,
  title={Back to Basics: Bank beta and Systemic Risk},
  author={Du, Ding, and Karolyi, Stephen, and Li, Feng},
  	Note = {Working Paper},
  year={2023}
}

@article{goldstein2005demand,
  title={Demand--deposit contracts and the probability of bank runs},
  author={Goldstein, Itay and Pauzner, Ady},
  journal={Journal of Finance},
  volume={60},
  number={3},
  pages={1293--1327},
  year={2005},
  publisher={Wiley Online Library}
}

@article{rochet2004coordination,
  title={Coordination failures and the lender of last resort: was {B}agehot right after all?},
  author={Rochet, Jean-Charles and Vives, Xavier},
  journal={Journal of the European Economic Association},
  volume={2},
  number={6},
  pages={1116--1147},
  year={2004},
  publisher={Oxford University Press}
}

@article{drechsler2017deposits,
  title={The deposits channel of monetary policy},
  author={Drechsler, Itamar and Savov, Alexi and Schnabl, Philipp},
  journal={Quarterly Journal of Economics},
  volume={132},
  number={4},
  pages={1819--1876},
  year={2017},
  publisher={Oxford University Press}
}

@article{drechsler2021banking,
  title={Banking on deposits: Maturity transformation without interest rate risk},
  author={Drechsler, Itamar and Savov, Alexi and Schnabl, Philipp},
  journal={Journal of Finance},
  volume={76},
  number={3},
  pages={1091--1143},
  year={2021},
  publisher={Wiley Online Library}
}

@article{BIS13,
	Author = {BCBS},
	Publisher = {Bank for International Settlements},
	Title = {Basel {III}: The liquidity coverage ratio and liquidity risk monitoring tools},
	Year = {2013}
}

@unpublished{jiang2023,
 Title = {Monetary Tightening and {U.S.} Bank Fragility in 2023: Mark-to-Market Losses and Uninsured Depositor Runs?},
 Author = {Jiang, Erica Xuewei and Matvos, Gregor and Piskorski, Tomasz and Seru, Amit},
Note = {Working Paper},
 Year = {2023}

}

@article{caglio2023,
  title={Flight to Safety in the Regional Bank Crisis of 2023},
  author={Caglio, Cecilia and Dlugosz, Jennifer and Rezende, Marcelo},
  journal={Available at SSRN 4457140},
  year={2023}
}

@article{metrick2024failure,
  title={The Failure of Silicon Valley Bank and the Panic of 2023},
  author={Metrick, Andrew},
  journal={Journal of Economic Perspectives},
  volume={38},
  number={1},
  pages={133--152},
  year={2024},
  publisher={American Economic Association 2014 Broadway, Suite 305, Nashville, TN 37203-2418}
}

@article{campbell1998econometrics,
  title={The econometrics of financial markets},
  author={Campbell, John Y and Lo, Andrew W and MacKinlay, A Craig and Whitelaw, Robert F},
  journal={Macroeconomic Dynamics},
  volume={2},
  number={4},
  pages={559--562},
  year={1998},
  publisher={Cambridge University Press}
}

@online{acharya2023svb,
  title={{SVB} and Beyond: The Banking Stress of 2023},
  author={Acharya, Viral V. and Richardson, Matthew P. and Schoenholtz, Kermit L. and Tuckman, Bruce and Berner, Richard and Cecchetti, Stephen G. and Kim, Sehwa and Kim, Seil and Philippon, Thomas and Ryan, Stephen G. and Savov, Alexi and Schnabl, Philipp and White, Lawrence J.},
  year={2023},
  month={7},
  day={17}
 
}

@article{OharaShaw1990,
  title={Deposit insurance and wealth effects: the value of being “too big to fail”},
  author={O'Hara, Maureen and Shaw, Wayne},
  journal={Journal of Finance},
  volume={45},
  number={5},
  pages={1587--1600},
  year={1990},
  publisher={Wiley Online Library}
}

@article{PenasUnal2004,
  title={Gains in bank mergers: Evidence from the bond markets},
  author={Penas, Mar{\i}a Fabiana and Unal, Haluk},
  journal={Journal of Financial Economics},
  volume={74},
  number={1},
  pages={149--179},
  year={2004},
  publisher={Elsevier}
}

@article{AllenGale1998,
	Author = {Allen, Franklin and Gale, Douglas},
		Issn = {1540-6261},
	Journal = {Journal of Finance},
	Number = {4},
	Pages = {1245--1284},
	Publisher = {Blackwell Publishers Inc.},
	Title = {Optimal Financial Crises},
	
	Volume = {53},
	Year = {1998}
	}

@article{CalomirisKahn1991,
	Author = {Calomiris, Charles W. and Kahn, Charles M.},
	Copyright = {Copyright {\copyright} 1991 American Economic Association},
	Issn = {00028282},
	Journal = {American Economic Review},
	Jstor_Articletype = {primary_article},
	Jstor_Formatteddate = {Jun., 1991},
	Number = {3},
	Pages = {497--513},
	Publisher = {American Economic Association},
	Title = {The Role of Demandable Debt in Structuring Optimal Banking Arrangements},
		Volume = {81},
	Year = {1991}}

@article{Diamond1983,
	Author = {Diamond, Douglas W. and Dybvig, Philip H.},
	Copyright = {Copyright {\copyright} 1983 The University of Chicago Press},
	Issn = {00223808},
	Journal = {Journal of Political Economy},
	Jstor_Articletype = {primary_article},
	Jstor_Formatteddate = {Jun., 1983},
	Number = {3},
	Pages = {401--419},
	Publisher = {The University of Chicago Press},
	Title = {Bank Runs, Deposit Insurance, and Liquidity},
	Volume = {91},
	Year = {1983}
	}

@article{DiamondRajan2001,
	Author = {Diamond, Douglas W. and Rajan, Raghuram G.},
	Copyright = {Copyright {\copyright} 2001 The University of Chicago Press},
	Issn = {00223808},
	Journal = {Journal of Political Economy},
	Jstor_Articletype = {primary_article},
	Jstor_Formatteddate = {Apr., 2001},
	Number = {2},
	Pages = {287--327},
	Publisher = {The University of Chicago Press},
	Review = {Three-period model with a somewhat complicated technology. Entrepreneur has project that requires him to work and can't precommit to working in a future period, therefore all contractual payments are subject to renegotiation. Lender has relation-ship specific knowledge that guarantees him a higher liquidation value than outside investors but he can also not commit to using this knowledge. To deterr the lender from renegotiating, a deposit contract creates fragility through the risk of runs in case the lender tries to renegotiate. This disciplining allows the lender to borrow from outside investors against the full value of his loan to the entrepreneur.},
	Title = {Liquidity Risk, Liquidity Creation, and Financial Fragility: A Theory of Banking},
	Volume = {109},
	Year = {2001}
	}

@article{EisenbachKeisterMcAndrewsYorulmazer2014,
	Author = {Eisenbach, Thomas M. and Todd Keister and James McAndrews and Tanju Yorulmazer},
	Journal = {Federal Reserve Bank of New York Economic Policy Review},
	Number = {1},
	Pages = {29-49},
	Title = {Stability of Funding Models: {A}n Analytical Framework},
	Volume = {20},
	Year = {2014}}

@article{BERGER1987501,
title = {Competitive viability in banking: Scale, scope, and product mix economies},
journal = {Journal of Monetary Economics},
volume = {20},
number = {3},
pages = {501-520},
year = {1987},
issn = {0304-3932},
author = {Allen N. Berger and Gerald A. Hanweck and David B. Humphrey}
}

@article{gale2020bank,
  title={Bank capital, fire sales, and the social value of deposits},
  author={Gale, D. and Yorulmazer, T.},
  journal={Econ Theory},
  volume={69},
  pages={919--963},
  year={2020},
  publisher={Springer}
}

@article{cookson2023,
  title={Social Media as a Bank Run Catalyst},
  author={Cookson, J Anthony and Fox, Corbin and Gil-Bazo, Javier and Imbet, Juan Felipe and Schiller, Christoph},
  journal={Available at SSRN 4422754},
  year={2023}
}

@article{AllenGale2000,
  title={Financial Contagion},
  author={Allen, Franklin and Gale, Douglas},
  journal={Journal of Political Economy},
  volume={108},
  number={1},
  pages={1--33},
  year={2000},
  publisher={The University of Chicago Press}
}

@incollection{CalomirisGorton1991,
  title={The origins of banking panics: models, facts, and bank regulation},
  author={Calomiris, Charles W and Gorton, Gary},
  booktitle={Financial Markets and Financial Crises},
  pages={109--174},
  year={1991},
  publisher={University of Chicago Press}
}

@article{Gorton1988,
  title={Banking panics and business cycles},
  author={Gorton, Gary},
  journal={Oxford Economic Papers},
  volume={40},
  number={4},
  pages={751--781},
  year={1988},
  publisher={Oxford University Press}
}

@article{ChariJagannathan1988,
  title={Banking panics, information, and rational expectations equilibrium},
  author={Chari, Varadarajan V and Jagannathan, Ravi},
  journal={Journal of Finance},
  volume={43},
  number={3},
  pages={749--761},
  year={1988},
  publisher={Wiley Online Library}
}

@article{GortonHuang2004,
  title={Liquidity, efficiency, and bank bailouts},
  author={Gorton, Gary and Huang, Lixin},
  journal={American Economic Review},
  volume={94},
  number={3},
  pages={455--483},
  year={2004},
  publisher={American Economic Association}
}

@article{chen199,
  title={Banking panics: The role of the first-come, first-served rule and information externalities},
  author={Chen, Yehning},
  journal={Journal of Political Economy},
  volume={107},
  number={5},
  pages={946--968},
  year={1999},
  publisher={The University of Chicago Press}
}

@article{AcharyaYorulmazer2008,
  title={Information contagion and bank herding},
  author={Acharya, Viral V and Yorulmazer, Tanju},
  journal={Journal of Money, Credit and Banking},
  volume={40},
  number={1},
  pages={215--231},
  year={2008},
  publisher={Wiley Online Library}
}

@article{cifuentes2005,
  title={Liquidity risk and contagion},
  author={Cifuentes, Rodrigo and Ferrucci, Gianluigi and Shin, Hyun Song},
  journal={Journal of the European Economic Association},
  volume={3},
  number={2-3},
  pages={556--566},
  year={2005},
  publisher={Oxford University Press}
}

@article{BootMilbournSchmeits2006,
  title={Credit ratings as coordination mechanisms},
  author={Boot, Arnoud WA and Milbourn, Todd T and Schmeits, Anjolein},
  journal={The Review of Financial Studies},
  volume={19},
  number={1},
  pages={81--118},
  year={2006},
  publisher={Oxford University Press}
}

@article{DercoleWagner2023,
  title={The green energy transition and the 2023 Banking Crisis},
  author={D’Ercole, Francesco and Wagner, Alexander F},
  journal={Finance Research Letters},
  pages={104493},
  year={2023},
  publisher={Elsevier}
}

@article{FlannerySorescu2023,
  title={Partial Effects of {Fed} Tightening on {US} Banks' Capital},
  author={Flannery, Mark J and Sorescu, Sorin M},
  journal={Available at SSRN 4424139},
  year={2023}
}

@article{DangGortonHolmstrom2020,
  title={The information view of financial crises},
  author={Dang, Tri Vi and Gorton, Gary and Holmstr{\"o}m, Bengt},
  journal={Annual Review of Financial Economics},
  volume={12},
  pages={39--65},
  year={2020},
  publisher={Annual Reviews}
}

@techreport{DrechslerSavovSchnablWang2023,
 title = "Banking on Uninsured Deposits",
 author = "Drechsler, Itamar and Savov, Alexi and Schnabl, Philipp and Wang, Olivier",
 institution = "National Bureau of Economic Research",
 type = "Working Paper",
 series = "Working Paper Series",
 number = "31138",
 year = "2023",
}

@article{HaddadHartmanMuir2023,
  title={Bank Fragility When Depositors Are the Asset},
  author={Haddad, Valentin and Hartman-Glaser, Barney and Muir, Tyler},
  journal={Available at SSRN 4412256},
  year={2023}
}

@article{jiang2023LimitedHedge,
  title={Limited Hedging and Gambling for Resurrection by {US} Banks During the 2022 Monetary Tightening?},
  author={Jiang, Erica Xuewei and Matvos, Gregor and Piskorski, Tomasz and Seru, Amit},
  journal={Available at SSRN},
  year={2023}
}

@article{GoldsmithYorulmazer2010,
  title={Liquidity, bank runs, and bailouts: spillover effects during the Northern Rock episode},
  author={Goldsmith-Pinkham, Paul and Yorulmazer, Tanju},
  journal={Journal of Financial Services Research},
  volume={37},
  pages={83--98},
  year={2010},
  publisher={Springer}
}

@techreport{Begenau2015,
 title = "Banks' Risk Exposures",
 author = "Begenau, Juliane and Piazzesi, Monika and Schneider, Martin",
 institution = "National Bureau of Economic Research",
 type = "Working Paper",
 series = "Working Paper Series",
 number = "21334",
 year = "2015",
}

@article{BenmelechYangZator2023,
  title={Bank Branch Density and Bank Runs},
  author={Benmelech, Efraim and Yang, Jun and Zator, Michal},
  year={2023},
  institution={National Bureau of Economic Research}
}

@article{LuckPlosserYounger2023,
  title={Bank Funding during the Current Monetary Policy Tightening Cycle},
  author={Luck, Stephan and Plosser, Matthew  and Josh Younger},
  journal={Liberty Street Economics},
  year={2023},
url = {https://libertystreeteconomics.newyorkfed.org/2023/05/bank-funding-during-the-current-monetary-policy-tightening-cycle/}
}

@article{Nelson2023,
  title={Silicon Valley Bank Would Have Passed The Liquidity Coverage Ratio Requirement},
  author={Bill Nelson},
  journal={Bank Policy Institute blog},
  year={2023},
url = {https://bpi.com/silicon-valley-bank-would-have-passed-the-liquidity-coverage-ratio-requirement/}
}

@article{KangLuckPlosser2023,
  title={Deposit Betas: Up, Up, and Away?},
  author={Kang-Landsberg, Alena and Luck, Stephan and Plosser, Matthew},
  journal={Liberty Street Economics},
  year={2023},
url = {https://libertystreeteconomics.newyorkfed.org/2023/04/deposit-betas-up-up-and-away/}
}

\clearpage
\setcounter{figure}{0} \renewcommand{\thefigure}{A\arabic{figure}}
\setcounter{table}{0} \renewcommand{\thetable}{A\arabic{table}}
\setcounter{equation}{0} \renewcommand{\theequation}{A\arabic{equation}}
\setcounter{section}{0} \renewcommand{\thesection}{A\Roman{section}}

\appendix

\begin{figure}[thbp]
    \centering
    \begin{subfigure}[b]{0.45\textwidth}
    \includegraphics[width=\linewidth]{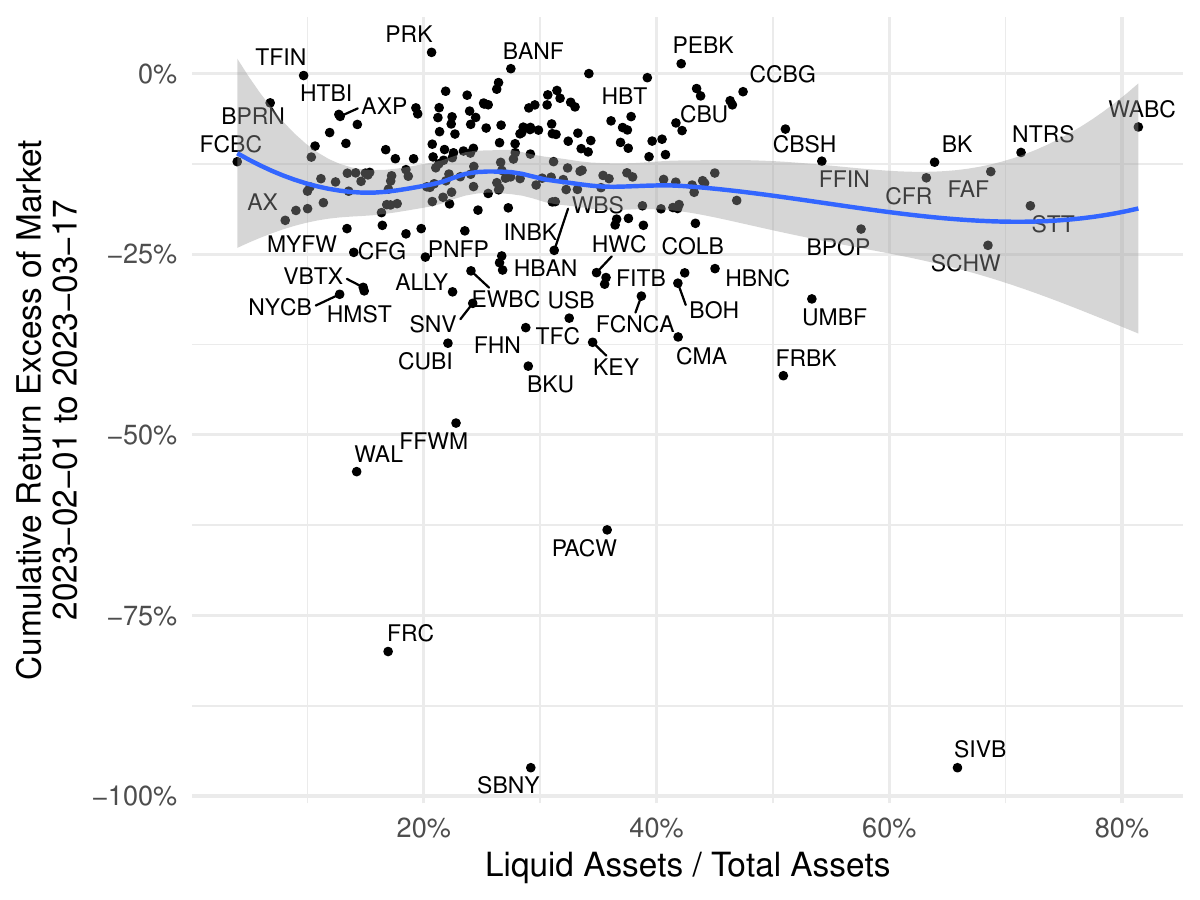}
        \caption{Liquid/ Assets}
    \end{subfigure}
     \begin{subfigure}[b]{0.45\textwidth}
        \includegraphics[width=\linewidth]{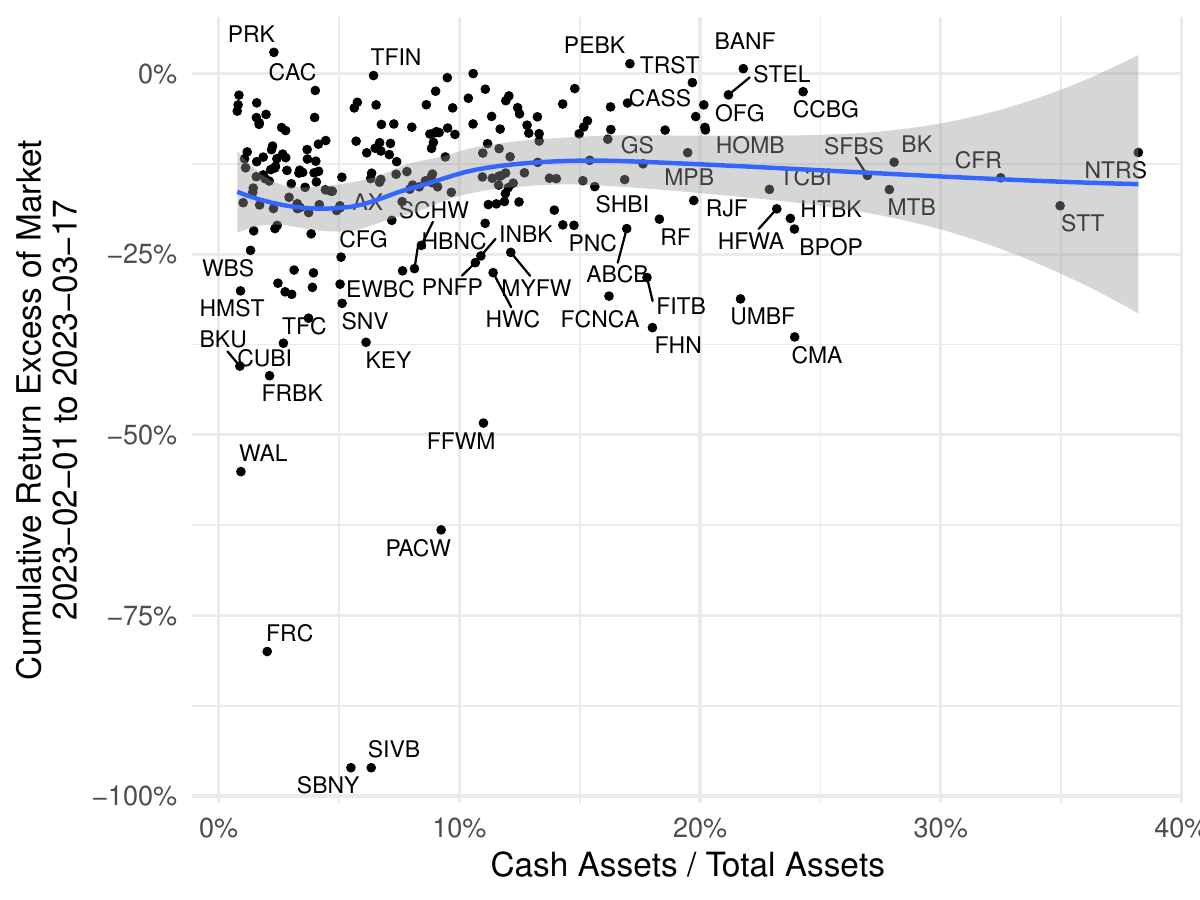}
        \caption{Cash/ Assets}
    \end{subfigure}
     \begin{subfigure}[b]{0.45\textwidth}
        \includegraphics[width=\linewidth]{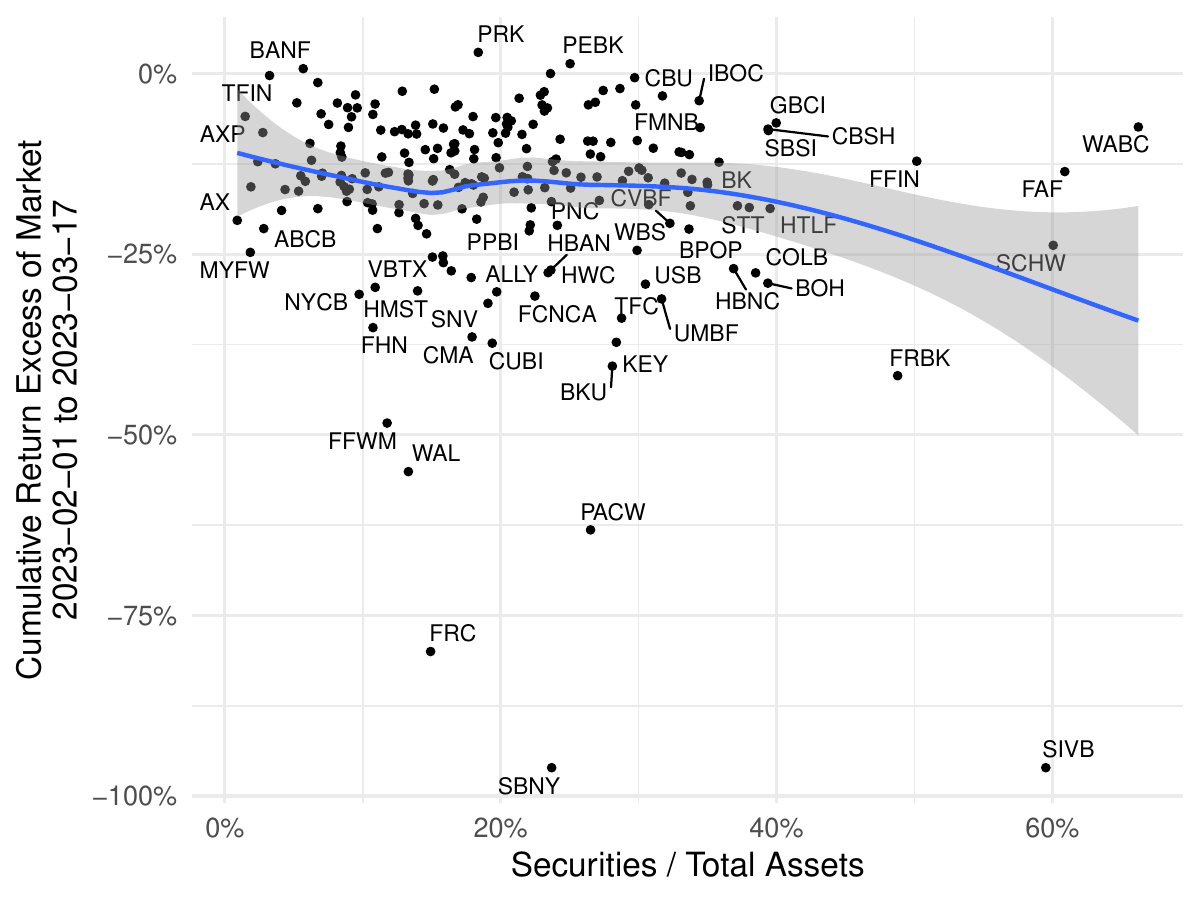} 
        \caption{Securities/ Assets}
    \end{subfigure}
    \caption{Early cumulative returns explained by liquid assets. Panel (a) plots the cumulative returns in excess of the S\&P 500 index for the banks in our sample from February 1, 2023 to March 17, 2023 against the liquid asset share (sum of liquid securities and cash scaled by total assets) measured in 2022q4. Panel (b) plots against cash scaled by total assets.  Panel (c) plots against liquid securities scaled by total assets. See Section \ref{sec:data} for details on the data construction.  The fitted lines are local linear regression (loess) curves. Stocks are denoted by their ticker symbol; Silicon Valley Bank's ticker is SIVB. }
    \label{fig:returns_v_liquid_assets}
\end{figure}

\clearpage

\begin{figure}[ht]
    \centering
    \includegraphics[width=0.75\linewidth]{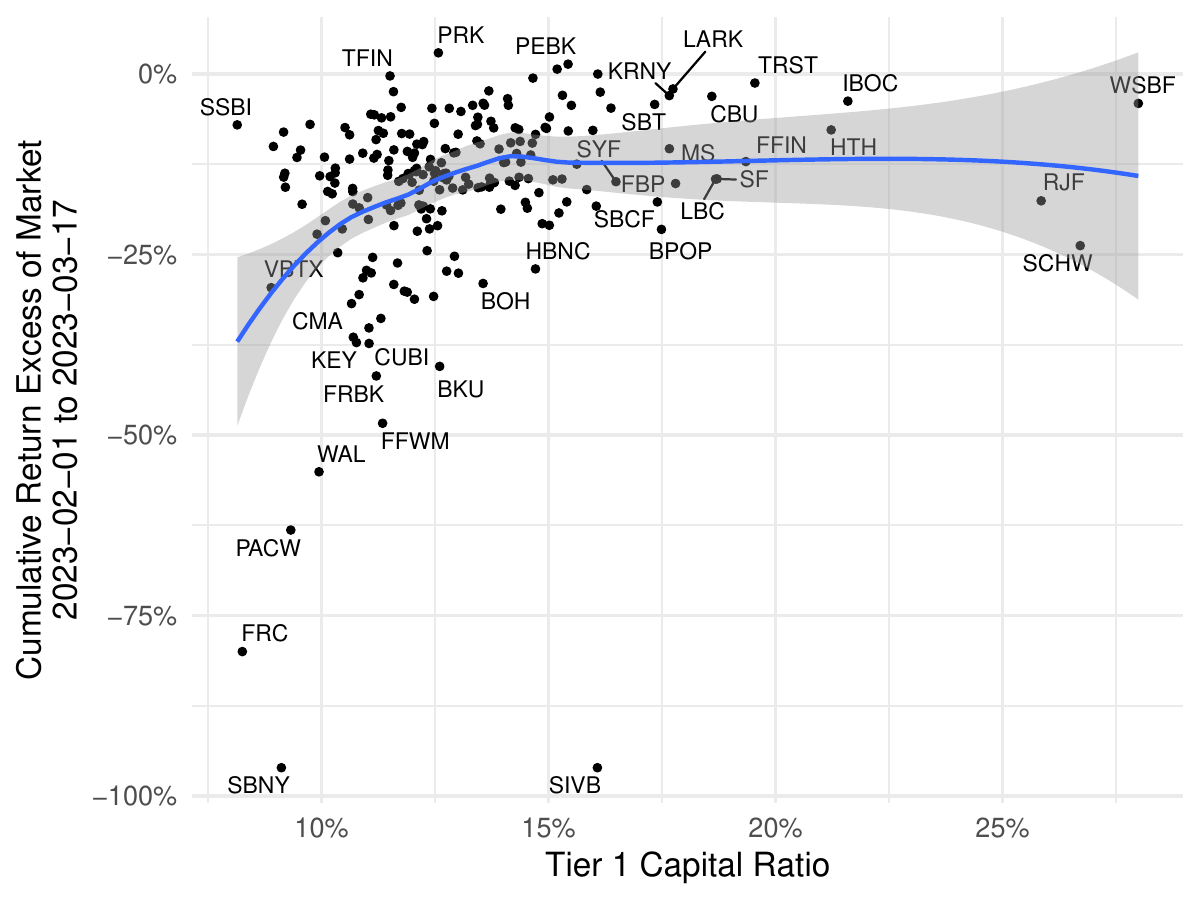}
    \caption{Early cumulative returns explained by Tier 1 Capital ratio. This figure plots the cumulative returns in excess of the S\&P 500 index for the banks in our sample from February 1, 2023 to March 17, 2023 against the Tier 1 capital ratio measured in 2022q4. See Section \ref{sec:data} for details on the data construction.  The fitted line is a local linear regression (loess) curve. Stocks are denoted by their ticker symbol; Silicon Valley Bank's ticker is SIVB. }
    \label{fig:returns_v_tier1capitalratio}
\end{figure}

\clearpage

\begin{figure}[thbp]
    \centering
    \includegraphics[width=0.65\linewidth]{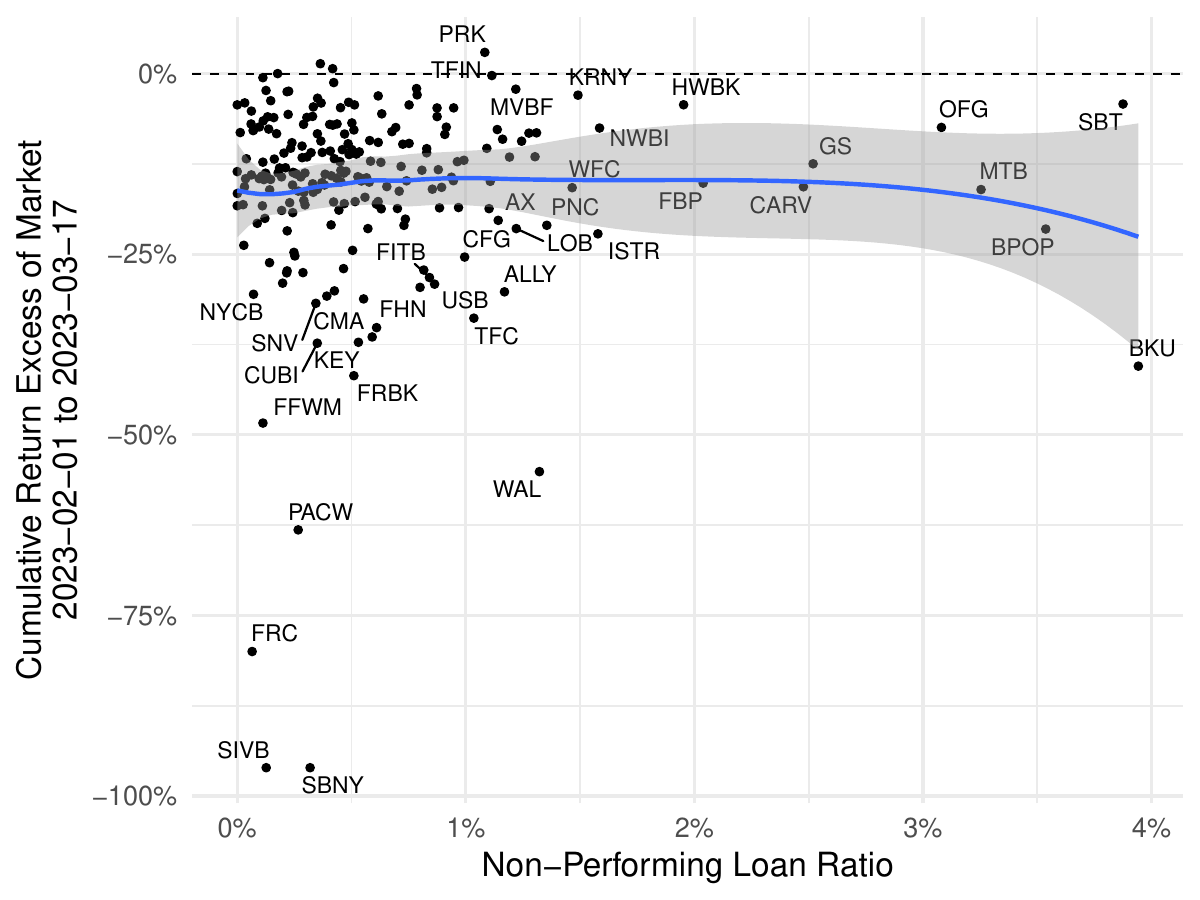}
    \caption{Early cumulative returns explained by non-performing loan ratio. This figure plots the cumulative returns in excess of the S\&P 500 index for the banks in our sample from February 1, 2023 to March 17, 2023 against the non-performing loan ratio (non-performing loans scaled by total loans) measured in 2022q4. See Section \ref{sec:data} for details on the data construction.  The fitted line is a local linear regression (loess) curve. Stocks are denoted by their ticker symbol; Silicon Valley Bank's ticker is SIVB. }
    \label{fig:returns_vs_NPL}
\end{figure}

\clearpage

\begin{table}[htbp]
   \caption{\label{tab:dep_htm_beta} Cumulative returns correlated with uninsured deposits and HTM, adjusted for beta. This table reports estimated coefficients of regressions with the cumulative returns in excess of the S\&P 500 index, adjusting for beta estimated in 2022, for the banks in our sample from February 1, 2023 to March 17, 2023 as the outcome.  In Column (1), we report the bivariate relationship with the uninsured deposit share (uninsured deposts as a share of total deposits) measured in 2022q4. Column (2) reports the coefficient with the hold-to-maturity asset share (hold-to-maturity assets as a share of total assets) measured in 2022q4. Column (3) reports unrealized hold-to-maturity losses scaled by tier 1 capital. Column (4) combines Columns (1)-(3). Column (5) adds the interaction of uninsured deposit share with HTM Asset Share and Unrealized HTM Losses. All variables (except for the cumulative returns) are mean zero and standarized to have standard deviation one, prior to interactions.}
   \bigskip
   \centering
   \begin{adjustbox}{width = 1.2\textwidth, center}
      \begin{tabular}{lccccc}
         \toprule
                                                                                  & (1)            & (2)            & (3)            & (4)            & (5)\\  
         \midrule 
         Constant                                                                 & -0.165$^{***}$ & -0.165$^{***}$ & -0.166$^{***}$ & -0.166$^{***}$ & -0.154$^{***}$\\   
                                                                                  & (0.008)        & (0.008)        & (0.008)        & (0.007)        & (0.007)\\   
         Uninsured Deposit Share                                                  & -0.048$^{***}$ &                &                & -0.034$^{***}$ & -0.029$^{***}$\\   
                                                                                  & (0.016)        &                &                & (0.011)        & (0.010)\\   
         HTM Asset Share                                                          &                & -0.046$^{**}$  &                & -0.020         & -0.007\\   
                                                                                  &                & (0.018)        &                & (0.016)        & (0.007)\\   
         Unrealized HTM Losses / Tier 1 Capital                                   &                &                & -0.063$^{***}$ & -0.049$^{***}$ & 0.007\\   
                                                                                  &                &                & (0.015)        & (0.009)        & (0.018)\\   
         Uninsured Deposit Share $\times$ HTM Asset Share                         &                &                &                &                & -0.009\\   
                                                                                  &                &                &                &                & (0.008)\\   
         Uninsured Deposit Share $\times$ Unrealized HTM Losses / Tier 1 Capital  &                &                &                &                & -0.042$^{***}$\\   
                                                                                  &                &                &                &                & (0.016)\\   
          \\
         Observations                                                             & 224            & 224            & 222            & 222            & 222\\  
         R$^2$                                                                    & 0.136          & 0.125          & 0.233          & 0.332          & 0.430\\  
         Adjusted R$^2$                                                           & 0.132          & 0.121          & 0.230          & 0.323          & 0.417\\  
         \bottomrule
      \end{tabular}
   \end{adjustbox}
\end{table}

\clearpage

\begin{table}[htbp]
   \caption{\label{tab:dep_htm_late} Cumulative returns correlated with uninsured deposits and HTM, long run. This table reports estimated coefficients of regressions with the cumulative returns in excess of the S\&P 500 index, adjusting for beta estimated in 2022, for the banks in our sample from February 1, 2023 to May 25, 2023 as the outcome.  In Column (1), we report the bivariate relationship with the uninsured deposit share (uninsured deposts as a share of total deposits) measured in 2022q4. Column (2) reports the coefficient with the hold-to-maturity asset share (hold-to-maturity assets as a share of total assets) measured in 2022q4. Column (3) reports unrealized hold-to-maturity losses scaled by tier 1 capital. Column (4) combines Columns (1)-(3). Column (5) adds the interaction of uninsured deposit share with HTM Asset Share and Unrealized HTM Losses. All variables (except for the cumulative returns) are mean zero and standarized to have standard deviation one, prior to interactions.}
   \bigskip
   \centering
   \begin{adjustbox}{width = 1.2\textwidth, center}
      \begin{tabular}{lccccc}
         \toprule
                                                                                  & (1)            & (2)            & (3)            & (4)            & (5)\\  
         \midrule 
         Constant                                                                 & -0.300$^{***}$ & -0.300$^{***}$ & -0.301$^{***}$ & -0.301$^{***}$ & -0.287$^{***}$\\   
                                                                                  & (0.010)        & (0.011)        & (0.010)        & (0.010)        & (0.010)\\   
         Uninsured Deposit Share                                                  & -0.053$^{***}$ &                &                & -0.045$^{***}$ & -0.041$^{***}$\\   
                                                                                  & (0.015)        &                &                & (0.012)        & (0.010)\\   
         HTM Asset Share                                                          &                & -0.035$^{*}$   &                & -0.007         & -0.0001\\   
                                                                                  &                & (0.018)        &                & (0.016)        & (0.010)\\   
         Unrealized HTM Losses / Tier 1 Capital                                   &                &                & -0.054$^{***}$ & -0.041$^{***}$ & 0.041$^{***}$\\   
                                                                                  &                &                & (0.013)        & (0.011)        & (0.016)\\   
         Uninsured Deposit Share $\times$ HTM Asset Share                         &                &                &                &                & -0.0004\\   
                                                                                  &                &                &                &                & (0.008)\\   
         Uninsured Deposit Share $\times$ Unrealized HTM Losses / Tier 1 Capital  &                &                &                &                & -0.063$^{***}$\\   
                                                                                  &                &                &                &                & (0.011)\\   
          \\
         Observations                                                             & 224            & 224            & 222            & 222            & 222\\  
         R$^2$                                                                    & 0.105          & 0.046          & 0.109          & 0.186          & 0.292\\  
         Adjusted R$^2$                                                           & 0.101          & 0.042          & 0.105          & 0.175          & 0.276\\  
         \bottomrule
      \end{tabular}
   \end{adjustbox}
\end{table}

\clearpage

\begin{table}[htbp]
   \caption{\label{tab:dep_htm_late_beta} Cumulative returns correlated with uninsured deposits and HTM, long run and adjusted for beta. This table reports estimated coefficients of regressions with the cumulative returns in excess of the S\&P 500 index for the banks in our sample from February 1, 2023 to May 25, 2023 as the outcome.  In Column (1), we report the bivariate relationship with the uninsured deposit share (uninsured deposts as a share of total deposits) measured in 2022q4. Column (2) reports the coefficient with the hold-to-maturity asset share (hold-to-maturity assets as a share of total assets) measured in 2022q4. Column (3) reports unrealized hold-to-maturity losses scaled by tier 1 capital. Column (4) combines Columns (1)-(3). Column (5) adds the interaction of uninsured deposit share with HTM Asset Share and Unrealized HTM Losses. All variables (except for the cumulative returns) are mean zero and standarized to have standard deviation one, prior to interactions.}
   \bigskip
   \centering
   \begin{adjustbox}{width = 1.2\textwidth, center}
      \begin{tabular}{lccccc}
         \toprule
                                                                                  & (1)            & (2)            & (3)            & (4)            & (5)\\  
         \midrule 
         Constant                                                                 & -0.283$^{***}$ & -0.283$^{***}$ & -0.284$^{***}$ & -0.284$^{***}$ & -0.270$^{***}$\\   
                                                                                  & (0.010)        & (0.011)        & (0.010)        & (0.010)        & (0.010)\\   
         Uninsured Deposit Share                                                  & -0.052$^{***}$ &                &                & -0.045$^{***}$ & -0.040$^{***}$\\   
                                                                                  & (0.015)        &                &                & (0.012)        & (0.010)\\   
         HTM Asset Share                                                          &                & -0.035$^{*}$   &                & -0.007         & -0.0001\\   
                                                                                  &                & (0.018)        &                & (0.015)        & (0.010)\\   
         Unrealized HTM Losses / Tier 1 Capital                                   &                &                & -0.054$^{***}$ & -0.041$^{***}$ & 0.041$^{***}$\\   
                                                                                  &                &                & (0.013)        & (0.011)        & (0.015)\\   
         Uninsured Deposit Share $\times$ HTM Asset Share                         &                &                &                &                & -0.0003\\   
                                                                                  &                &                &                &                & (0.008)\\   
         Uninsured Deposit Share $\times$ Unrealized HTM Losses / Tier 1 Capital  &                &                &                &                & -0.063$^{***}$\\   
                                                                                  &                &                &                &                & (0.011)\\   
          \\
         Observations                                                             & 224            & 224            & 222            & 222            & 222\\  
         R$^2$                                                                    & 0.105          & 0.046          & 0.110          & 0.187          & 0.294\\  
         Adjusted R$^2$                                                           & 0.101          & 0.042          & 0.106          & 0.175          & 0.277\\  
         \bottomrule
      \end{tabular}
   \end{adjustbox}
\end{table}

\clearpage

\begin{table}[htbp]
   \caption{\label{tab:liquid_assets_beta} Cumulative returns correlated with liquid assets, adjusted for beta. This table reports estimated coefficients of regressions with the cumulative returns in excess of the S\&P 500 index, adjusted for beta estimated in 2002, for the banks in our sample from February 1, 2023 to March 17, 2023 as the outcome. 
      In Column (1), we report the bivariate relationship with the liquid asset share (securities + cash scaled by total
   assets) measured in 2022q4. 
      Column (2) adds uninsured deposit share to Column (1).
      Column (3) reports the coefficient with the cash scaled by total assets and securities scaled by total assets. 
       Column (4) adds uninsured deposit share to Column (3). 
       Column (5) interacts uninsured deposit share with cash and securities. 
       All variables (except for the cumulative returns) are mean zero and standarized to have standard deviation one, prior to interactions.}
   \bigskip
   \centering
   \begin{adjustbox}{width = 1.2\textwidth, center}
      \begin{tabular}{lccccc}
         \toprule
                                                                     & (1)            & (2)            & (3)            & (4)            & (5)\\  
         \midrule 
         Constant                                                    & -0.165$^{***}$ & -0.165$^{***}$ & -0.165$^{***}$ & -0.165$^{***}$ & -0.167$^{***}$\\   
                                                                     & (0.009)        & (0.009)        & (0.008)        & (0.008)        & (0.008)\\   
         Liquid Assets / Total Assets                                & -0.009         &                & 0.008          &                &   \\   
                                                                     & (0.012)        &                & (0.011)        &                &   \\   
         Cash / Total Assets                                         &                & 0.017$^{**}$   &                & 0.033$^{***}$  & 0.024$^{***}$\\   
                                                                     &                & (0.007)        &                & (0.010)        & (0.008)\\   
         Securities / Total Assets                                   &                & -0.021         &                & -0.009         & -0.005\\   
                                                                     &                & (0.014)        &                & (0.011)        & (0.008)\\   
         Uninsured Deposit Share                                     &                &                & -0.051$^{***}$ & -0.056$^{***}$ & -0.054$^{***}$\\   
                                                                     &                &                & (0.016)        & (0.016)        & (0.014)\\   
         Cash / Total Assets $\times$ Uninsured Deposit Share        &                &                &                &                & 0.016$^{**}$\\   
                                                                     &                &                &                &                & (0.007)\\   
         Securities / Total Assets $\times$ Uninsured Deposit Share  &                &                &                &                & -0.014\\   
                                                                     &                &                &                &                & (0.016)\\   
          \\
         Observations                                                & 224            & 224            & 224            & 224            & 224\\  
         R$^2$                                                       & 0.005          & 0.043          & 0.139          & 0.203          & 0.244\\  
         Adjusted R$^2$                                              & 0.0002         & 0.034          & 0.131          & 0.192          & 0.226\\  
         \bottomrule
      \end{tabular}
   \end{adjustbox}
\end{table}

\clearpage

\begin{table}[htbp]
   \caption{\label{tab:liquid_assets_late} Cumulative returns correlated with liquid assets, long run. This table reports estimated coefficients of regressions with the cumulative returns in excess of the S\&P 500 index for the banks in our sample from February 1, 2023 to March 17, 2023 as the outcome. 
      In Column (1), we report the bivariate relationship with the liquid asset share (securities + cash scaled by total
   assets) measured in 2022q4. 
      Column (2) adds uninsured deposit share to Column (1). 
      Column (3) reports the coefficient with the cash scaled by total assets and securities scaled by total assets. 
       Column (4) adds uninsured deposit share to Column (3). 
       Column (5) interacts uninsured deposit share with cash and securities. 
       All variables (except for the cumulative returns) are mean zero and standarized to have standard deviation one, prior to interactions.}
   \bigskip
   \centering
   \begin{adjustbox}{width = 1.2\textwidth, center}
      \begin{tabular}{lccccc}
         \toprule
                                                                     & (1)            & (2)            & (3)            & (4)            & (5)\\  
         \midrule 
         Constant                                                    & -0.300$^{***}$ & -0.300$^{***}$ & -0.300$^{***}$ & -0.300$^{***}$ & -0.301$^{***}$\\   
                                                                     & (0.011)        & (0.011)        & (0.010)        & (0.010)        & (0.010)\\   
         Liquid Assets / Total Assets                                & 0.004          &                & 0.024$^{**}$   &                &   \\   
                                                                     & (0.013)        &                & (0.012)        &                &   \\   
         Cash / Total Assets                                         &                & 0.033$^{***}$  &                & 0.053$^{***}$  & 0.048$^{***}$\\   
                                                                     &                & (0.010)        &                & (0.011)        & (0.012)\\   
         Securities / Total Assets                                   &                & -0.015         &                & -0.0003        & -0.001\\   
                                                                     &                & (0.013)        &                & (0.011)        & (0.008)\\   
         Uninsured Deposit Share                                     &                &                & -0.060$^{***}$ & -0.068$^{***}$ & -0.069$^{***}$\\   
                                                                     &                &                & (0.015)        & (0.015)        & (0.014)\\   
         Cash / Total Assets $\times$ Uninsured Deposit Share        &                &                &                &                & 0.007\\   
                                                                     &                &                &                &                & (0.008)\\   
         Securities / Total Assets $\times$ Uninsured Deposit Share  &                &                &                &                & -0.0004\\   
                                                                     &                &                &                &                & (0.016)\\   
          \\
         Observations                                                & 224            & 224            & 224            & 224            & 224\\  
         R$^2$                                                       & 0.0006         & 0.050          & 0.124          & 0.201          & 0.204\\  
         Adjusted R$^2$                                              & -0.004         & 0.041          & 0.116          & 0.190          & 0.186\\  
         \bottomrule
      \end{tabular}
   \end{adjustbox}
\end{table}

\clearpage

\begin{table}[htbp]
   \caption{\label{tab:liquid_assets_late_beta} Cumulative returns correlated with liquid assets, adjusted for beta, long run. This table reports estimated coefficients of regressions with the cumulative returns in excess of the S\&P 500 index, adjusted for beta estimated in 2002, for the banks in our sample from February 1, 2023 to March 17, 2023 as the outcome.
      In Column (1), we report the bivariate relationship with the liquid asset share (securities + cash scaled by total
   assets) measured in 2022q4. 
      Column (2) adds uninsured deposit share to Column (1). 
      Column (3) reports the coefficient with the cash scaled by total assets and securities scaled by total assets. 
       Column (4) adds uninsured deposit share to Column (3). 
       Column (5) interacts uninsured deposit share with cash and securities. 
       All variables (except for the cumulative returns) are mean zero and standarized to have standard deviation one, prior to interactions.}
   \bigskip
   \centering
   \begin{adjustbox}{width = 1.2\textwidth, center}
      \begin{tabular}{lccccc}
         \toprule
                                                                     & (1)            & (2)            & (3)            & (4)            & (5)\\  
         \midrule 
         Constant                                                    & -0.283$^{***}$ & -0.283$^{***}$ & -0.283$^{***}$ & -0.283$^{***}$ & -0.285$^{***}$\\   
                                                                     & (0.011)        & (0.011)        & (0.010)        & (0.010)        & (0.010)\\   
         Liquid Assets / Total Assets                                & 0.004          &                & 0.024$^{**}$   &                &   \\   
                                                                     & (0.013)        &                & (0.012)        &                &   \\   
         Cash / Total Assets                                         &                & 0.032$^{***}$  &                & 0.052$^{***}$  & 0.048$^{***}$\\   
                                                                     &                & (0.010)        &                & (0.011)        & (0.012)\\   
         Securities / Total Assets                                   &                & -0.014         &                & -0.0002        & -0.0009\\   
                                                                     &                & (0.013)        &                & (0.011)        & (0.008)\\   
         Uninsured Deposit Share                                     &                &                & -0.060$^{***}$ & -0.067$^{***}$ & -0.068$^{***}$\\   
                                                                     &                &                & (0.015)        & (0.015)        & (0.014)\\   
         Cash / Total Assets $\times$ Uninsured Deposit Share        &                &                &                &                & 0.007\\   
                                                                     &                &                &                &                & (0.008)\\   
         Securities / Total Assets $\times$ Uninsured Deposit Share  &                &                &                &                & -0.0005\\   
                                                                     &                &                &                &                & (0.016)\\   
          \\
         Observations                                                & 224            & 224            & 224            & 224            & 224\\  
         R$^2$                                                       & 0.0006         & 0.049          & 0.124          & 0.200          & 0.203\\  
         Adjusted R$^2$                                              & -0.004         & 0.041          & 0.116          & 0.189          & 0.185\\  
         \bottomrule
      \end{tabular}
   \end{adjustbox}
\end{table}

\clearpage

\begin{table}[htbp]
   \caption{\label{tab:npl_capital_beta} Cumulative returns correlated with NPL and Tier 1 capital, adjusted for beta. This table reports estimated coefficients of regressions with the cumulative returns in excess of the S\&P 500 index, adjusted for beta estimated in 2002, for the banks in our sample from February 1, 2023 to March 17, 2023 as the outcome. 
      In Column (1), we report the bivariate relationship with 
      the tier 1 capital ratio measured in 2022q4. %
      Column (2) reports the coefficient with the non-performing loan ratio (non-performing loans scaled by total loans) measured in 2022q4.
      Column (3) combines Column (1) and (2). 
       Column (4) adds uninsured deposit share to Column (3). 
       Column (5) interacts uninsured deposit share with non-performing loans and tier 1 capital. 
       All variables (except for the cumulative returns) are mean zero and standarized to have standard deviation one, prior to interactions.}
   \bigskip
   \centering
   \begin{adjustbox}{width = 1.2\textwidth, center}
      \begin{tabular}{lccccc}
         \toprule
                                                                              & (1)            & (2)            & (3)            & (4)            & (5)\\  
         \midrule 
         Constant                                                             & -0.168$^{***}$ & -0.165$^{***}$ & -0.168$^{***}$ & -0.167$^{***}$ & -0.160$^{***}$\\   
                                                                              & (0.009)        & (0.009)        & (0.009)        & (0.008)        & (0.009)\\   
         Tier 1 Capital Ratio                                                 & 0.033$^{***}$  &                & 0.033$^{***}$  & 0.023$^{**}$   & 0.038$^{*}$\\   
                                                                              & (0.011)        &                & (0.011)        & (0.011)        & (0.020)\\   
         Non-Performing Loans / Total Loans                                   &                & 0.004          & 0.002          & 0.001          & -0.002\\   
                                                                              &                & (0.009)        & (0.009)        & (0.008)        & (0.008)\\   
         Uninsured Deposit Share                                              &                &                &                & -0.046$^{***}$ & -0.049$^{***}$\\   
                                                                              &                &                &                & (0.018)        & (0.017)\\   
         Tier 1 Capital Ratio $\times$ Uninsured Deposit Share                &                &                &                &                & 0.033\\   
                                                                              &                &                &                &                & (0.025)\\   
         Non-Performing Loans / Total Loans $\times$ Uninsured Deposit Share  &                &                &                &                & 0.002\\   
                                                                              &                &                &                &                & (0.012)\\   
          \\
         Observations                                                         & 216            & 224            & 216            & 216            & 216\\  
         R$^2$                                                                & 0.062          & 0.0009         & 0.063          & 0.171          & 0.222\\  
         Adjusted R$^2$                                                       & 0.058          & -0.004         & 0.054          & 0.159          & 0.203\\  
         \bottomrule
      \end{tabular}
   \end{adjustbox}
\end{table}

\clearpage

\begin{table}[htbp]
   \caption{\label{tab:npl_capital_late} Cumulative returns correlated with NPL and Tier 1 capital, long run. This table reports estimated coefficients of regressions with the cumulative returns in excess of the S\&P 500 index for the banks in our sample from February 1, 2023 to May 25, 2023 as the outcome. 
      In Column (1), we report the bivariate relationship with 
      the tier 1 capital ratio measured in 2022q4. %
      Column (2) reports the coefficient with the non-performing loan ratio (non-performing loans scaled by total loans) measured in 2022q4.
      Column (3) combines Column (1) and (2). 
       Column (4) adds uninsured deposit share to Column (3). 
       Column (5) interacts uninsured deposit share with non-performing loans and tier 1 capital. 
       All variables (except for the cumulative returns) are mean zero and standarized to have standard deviation one, prior to interactions.}
   \bigskip
   \centering
   \begin{adjustbox}{width = 1.2\textwidth, center}
      \begin{tabular}{lccccc}
         \toprule
                                                                              & (1)            & (2)            & (3)            & (4)            & (5)\\  
         \midrule 
         Constant                                                             & -0.302$^{***}$ & -0.300$^{***}$ & -0.302$^{***}$ & -0.301$^{***}$ & -0.294$^{***}$\\   
                                                                              & (0.011)        & (0.011)        & (0.011)        & (0.010)        & (0.011)\\   
         Tier 1 Capital Ratio                                                 & 0.041$^{***}$  &                & 0.040$^{***}$  & 0.029$^{***}$  & 0.043$^{**}$\\   
                                                                              & (0.011)        &                & (0.011)        & (0.011)        & (0.018)\\   
         Non-Performing Loans / Total Loans                                   &                & 0.017$^{*}$    & 0.016          & 0.015          & 0.010\\   
                                                                              &                & (0.010)        & (0.010)        & (0.009)        & (0.009)\\   
         Uninsured Deposit Share                                              &                &                &                & -0.051$^{***}$ & -0.052$^{***}$\\   
                                                                              &                &                &                & (0.017)        & (0.016)\\   
         Tier 1 Capital Ratio $\times$ Uninsured Deposit Share                &                &                &                &                & 0.032\\   
                                                                              &                &                &                &                & (0.021)\\   
         Non-Performing Loans / Total Loans $\times$ Uninsured Deposit Share  &                &                &                &                & 0.011\\   
                                                                              &                &                &                &                & (0.013)\\   
          \\
         Observations                                                         & 216            & 224            & 216            & 216            & 216\\  
         R$^2$                                                                & 0.063          & 0.011          & 0.072          & 0.156          & 0.190\\  
         Adjusted R$^2$                                                       & 0.058          & 0.007          & 0.063          & 0.144          & 0.171\\  
         \bottomrule
      \end{tabular}
   \end{adjustbox}
\end{table}

\clearpage

\begin{table}[htbp]
   \caption{\label{tab:npl_capital_late_beta} Cumulative returns correlated with NPL and Tier 1 capital, adjusted for beta, long run. This table reports estimated coefficients of regressions with the cumulative returns in excess of the S\&P 500 index, adjusted for beta estimated in 2002, for the banks in our sample from February 1, 2023 to May 25, 2023 as the outcome. 
      In Column (1), we report the bivariate relationship with 
      the tier 1 capital ratio measured in 2022q4. %
      Column (2) reports the coefficient with the non-performing loan ratio (non-performing loans scaled by total loans) measured in 2022q4.
      Column (3) combines Column (1) and (2).
       Column (4) adds uninsured deposit share to Column (3). 
       Column (5) interacts uninsured deposit share with non-performing loans and tier 1 capital. 
       All variables (except for the cumulative returns) are mean zero and standarized to have standard deviation one, prior to interactions.}
   \bigskip
   \centering
   \begin{adjustbox}{width = 1.2\textwidth, center}
      \begin{tabular}{lccccc}
         \toprule
                                                                              & (1)            & (2)            & (3)            & (4)            & (5)\\  
         \midrule 
         Constant                                                             & -0.285$^{***}$ & -0.283$^{***}$ & -0.285$^{***}$ & -0.284$^{***}$ & -0.277$^{***}$\\   
                                                                              & (0.011)        & (0.011)        & (0.011)        & (0.010)        & (0.011)\\   
         Tier 1 Capital Ratio                                                 & 0.041$^{***}$  &                & 0.040$^{***}$  & 0.029$^{***}$  & 0.043$^{**}$\\   
                                                                              & (0.011)        &                & (0.011)        & (0.011)        & (0.018)\\   
         Non-Performing Loans / Total Loans                                   &                & 0.017$^{*}$    & 0.015          & 0.014          & 0.010\\   
                                                                              &                & (0.010)        & (0.010)        & (0.009)        & (0.009)\\   
         Uninsured Deposit Share                                              &                &                &                & -0.051$^{***}$ & -0.051$^{***}$\\   
                                                                              &                &                &                & (0.016)        & (0.015)\\   
         Tier 1 Capital Ratio $\times$ Uninsured Deposit Share                &                &                &                &                & 0.032\\   
                                                                              &                &                &                &                & (0.021)\\   
         Non-Performing Loans / Total Loans $\times$ Uninsured Deposit Share  &                &                &                &                & 0.011\\   
                                                                              &                &                &                &                & (0.013)\\   
          \\
         Observations                                                         & 216            & 224            & 216            & 216            & 216\\  
         R$^2$                                                                & 0.062          & 0.011          & 0.071          & 0.155          & 0.190\\  
         Adjusted R$^2$                                                       & 0.058          & 0.007          & 0.062          & 0.143          & 0.170\\  
         \bottomrule
      \end{tabular}
   \end{adjustbox}
\end{table}

\end{document}